\documentclass[aps,prd,nofootinbib,showkeys,preprint,floatfix]{revtex4}

\usepackage{amssymb}
\usepackage{amsmath}
\usepackage{amsfonts}
\usepackage{graphicx}
\usepackage{color}
\usepackage{xspace}
\usepackage{ulem}
\usepackage{lscape}
\usepackage{ wasysym }

\usepackage{float}
\usepackage{subfig}
\usepackage{mathrsfs}
\usepackage{slashed}

\usepackage{multirow,rotating}

\def\gsim{\raise0.3ex\hbox{$\;>$\kern-0.75em\raise-1.1ex\hbox{$\sim\;$}}}
\def\lsim{\raise0.3ex\hbox{$\;<$\kern-0.75em\raise-1.1ex\hbox{$\sim\;$}}}

\newcommand{\ba}[1]{\begin{eqnarray} \label{(#1)}}
\newcommand{\ea}{\end{eqnarray}}

\newcommand{\FIG}[1]{fig.~\ref{#1}}

\newcommand{\AddrAHEP}{
  {\it AHEP Group, Instituto de F\'{\i}sica Corpuscular --
    C.S.I.C./Universitat de Val{\`e}ncia \\
    Edificio de Institutos de Paterna, Apartado 22085,
  E--46071 Val{\`e}ncia, Spain}}
  
  \newcommand{\AddrUFSM}{
Departamento de F\' isica, Facultad de Ciencias, Universidad de La Serena, \\
Avenida Cisternas 1200, La Serena, Chile.  \\
Centro-Cient\'\i fico-Tecnol\'{o}gico de Valpara\'\i so, \\ 
Casilla 110-V, Valpara\'\i so,  Chile.
}

\def\gsim{\raise0.3ex\hbox{$\;>$\kern-0.75em\raise-1.1ex\hbox{$\sim\;$}}}
\def\lsim{\raise0.3ex\hbox{$\;<$\kern-0.75em\raise-1.1ex\hbox{$\sim\;$}}}

\begin{document}

\preprint{IFIC/17-21}

\title{Loop neutrino masses from $d=7$ operator}

\author{R. Cepedello}\email{ricepe@ific.uv.es}\affiliation{\AddrAHEP}
\author{J.C. Helo} \email{juancarlos.helo@usm.cl}\affiliation{\AddrUFSM}
\author{M. Hirsch} \email{mahirsch@ific.uv.es}\affiliation{\AddrAHEP}



\vskip10mm
\begin{abstract}

We discuss the generation of small neutrino masses from $d=7$ 1-loop
diagrams. We first systematically analyze all possible $d=7$ 1-loop
topologies. There is a total of 48 topologies, but only 8 of these can
lead to ``genuine'' $d=7$ neutrino masses. Here, we define genuine
models to be models in which neither $d=5$ nor $d=7$ tree-level masses
nor a $d=5$ 1-loop mass appear, such that the $d=7$ 1-loop is the
leading order contribution to the neutrino masses. All genuine models
can then be organized w.r.t. their particle content. We find there is
only one diagram with no representation larger than triplet, while
there are 22 diagrams with quadruplets. We briefly discuss three
minimal example models of this kind.

\end{abstract}

\maketitle

%
\section{Introduction\label{Sect:Intro}}

\noindent
If neutrinos are Majorana particles, one can roughly estimate their
mass as:
\begin{equation}\label{eq:mnu}
m_{\nu} \propto \frac{v^2}{\Lambda} 
               \times \Big(\frac{1}{16 \pi^2}\Big)^n 
               \times \epsilon 
               \times \Big(\frac{v}{\Lambda}\Big)^{d-5} .
\end{equation}
Here, $\Lambda$ is the energy scale of new physics, where lepton
number violation (LNV) occurs and $v$ is the standard model vacuum
expectation value. The different terms in eq. (\ref{eq:mnu}) can be
understood easily. The first term corresponds to the famous Weinberg
operator, ${\cal O}^W \equiv {\cal O}^{d=5} \propto LLHH$
\cite{Weinberg:1979sa}.  This operator can be generated either at
tree-level or at loop level. The second term in eq. (\ref{eq:mnu})
takes into account this fact, with $n=0,1,2,\cdots$ being the number
of loops. Then, there are models of neutrino mass, in which the
Weinberg operator is suppressed by some small factor $\epsilon$, which
could be either due to some small coupling in the corresponding model
or due to some nearly conserved symmetry.  R-parity violating
supersymmetry is an example of the former
\cite{Dreiner:1997uz,Hirsch:2000ef}, models such as the inverse
\cite{Mohapatra:1986bd} or the linear
\cite{Akhmedov:1995ip,Akhmedov:1995vm} seesaw are examples of the
latter. And, finally, neutrino masses could be due to higher
dimensional operators. This is expressed by the last term in
eq. (\ref{eq:mnu}), with $d=5,7, \cdots$ the dimension of the
operator.

In this paper we will focus on $d=7$ neutrino mass models at 1-loop
order. Our aim is to give a systematic analysis of such models,
constructing first all possible 1-loop topologies and then identify
those topologies, which {\em allow to construct genuine} models.
Here, we define {\em genuine} models such models, where the 1-loop
$d=7$ contribution to the neutrino mass gives the leading order
contribution. This assumption implies, of course, that both the $d=5$
and the $d=7$ tree-level, as well as the $d=5$ 1-loop, contributions
should be absent.

$d=5$ neutrino masses have been studied extensively in the literature.
In \cite{Ma:1998dn} it was shown that there are only three tree-level
realizations of ${\cal O}^W$ at tree-level. A systematic analysis
of the Weinberg operator at 1-loop level was presented in
\cite{Bonnet:2012kz}, at 2-loop level in \cite{Sierra:2014rxa}, see
also \cite{Farzan:2012ev} for a general discussion of tree versus
loop neutrino masses.

Disregarding derivative operators, the authors of \cite{Babu:2001ex}
have written down all $\Delta L=2$ operators up to $d=11$. 
Only one of these operators is important for us here:
\begin{equation}\label{eq:defo7}
{\cal O}^{d=7} \propto LLHHHH^{\dagger}
\end{equation}
All other $d=7$ operators in the list of \cite{Babu:2001ex} will lead
to $d=5$ 1-loop neutrino mass models, while the $d=9$ and $d=11$
operators can lead only to $d=7$ neutrino masses, if the underlying
model is 2-loop or higher. Note that $\Delta L=2$ operators with
derivatives have been studied in \cite{delAguila:2012nu}. Two
operators with derivatives at $d=7$ exist, but neither can 
lead to a 1-loop $d=7$ model, see also the discussion in
\cite{Helo:2016vsi}.

Bonnet et al. \cite{Bonnet:2009ej} analyzed the $d=7$ operator of
eq. (\ref{eq:defo7}) at tree-level in detail. As noted in this work,
the $d=7$ operator of eq. (\ref{eq:defo7}) will always also generate a
higher order $d=5$ neutrino mass:
\begin{equation}\label{eq:nlp1}
\frac{1}{\Lambda^3}LLHHHH^{\dagger} \rightarrow \frac{1}{16 \pi^2} 
\frac{1}{\Lambda}LLHH
\end{equation}
One can straightforwardly estimate that this loop contribution will
become more important than the tree-level one if $(\Lambda/v) \gsim 4
\pi$. This means $\Lambda \lsim 2$ TeV is required for the $d=7$
contribution to dominate.  Since this is unavoidable in the standard
model (SM), the authors of \cite{Bonnet:2009ej} considered a two Higgs
doublet extension of the SM in their discussion of the $d=7$
tree-level neutrino mass. \footnote{$HH^{\dagger}$ is a singlet under
  any discrete symmetry. With more than one Higgs it is possible to
  introduce an additional discrete symmetry, under which the two
  Higgses transform differently.} We instead will stick to only the SM
Higgs and take eq. (\ref{eq:nlp1}) as a motivation that any $d=7$
model of neutrino mass {\em must have new particles below 2 TeV},
otherwise it will not give the leading contribution to the neutrino
mass matrix.

As mentioned above, both $d=5$ and $d=7$ tree-level contributions
should be forbidden, otherwise the $d=7$ 1-loop contribution might be
just some minor correction to the neutrino mass matrix. Absence of
these lower order contributions could be attributed to either: (i) the
existence of some symmetry; or (ii) absence of fields which generate
neutrino masses at lower order. An example of the former at $d=5$ is
the scotogenic model \cite{Ma:2006km}. In this model, a right-handed
neutrino (plus an extra doublet scalar) exists, but due to a $Z_2$
symmetry, there is no tree-level $d=5$ neutrino mass.\footnote{The
  well-known bonus of the $Z_2$ symmetry is that it allows to
  ``stabilize'' the lightest $Z_2$ odd particle, thus relating the
  stability of the dark matter to the generation of neutrino masses.}
Instead neutrinos have mass at ($d=5$) 1-loop order. The classic
example for (ii) is the Zee model \cite{Zee:1980ai}. In the Zee model,
none of the particles necessary for a tree-level seesaw exist. Instead
an additional charged singlet scalar (plus an additional doublet
scalar) generate neutrino masses radiatively.

We will not discuss in detail additional discrete (or gauge)
symmetries here, since models of $d=7$ neutrino masses have been
discussed in this context already in a number of references, see for
example
\cite{Bonnet:2009ej,Chen:2006hn,Kanemura:2010bq,Krauss:2011ur,Krauss:2013gy}.
Also we will not consider $d=7$ operators with additional singlets,
see for example \cite{Gogoladze:2008wz,Gu:2009hu}.  Instead, we will
follow the second route mentioned above: After constructing all 1-loop
$d=7$ topologies, we will classify the underlying models according to
their particle content.

The rest of this paper is organized as follows. In the next section,
we will first give a short summary of neutrino masses at tree-level at
$d=5$ and $d=7$, as well as $d=5$ at 1-loop order.  This provides the
basis for the discussion in section \ref{Sect:Class}. Section
\ref{Sect:Class} then provides the core of our present work. It
discusses all possible topologies and classifies them into different
groups. The section then also introduces the three most minimal
example models that one can construct at $d=7$ 1-loop order. We then
close with a short summary. More complete lists of topologies and
diagrams are relegated to the appendix.

\section{Preliminaries\label{Sect:Prelim}}

Since we are interested in identifying models, in which a 1-loop $d=7$
diagram gives the leading order contribution to the neutrino mass
matrix, we first need to discuss briefly neutrino masses at lower
order.  When discussing possible models will use $S$ (and $\phi$) for
scalars and $\psi$ ($\chi$) for fermions.  For a more compact notation
we will also use a notation which gives the $SU(2)_L$ representation
and hypercharge in the form ${\bf R}_Y$ with a superscript $S$ or $F$,
where necessary, i.e. for example ${\bf 5}_1^S$ is a scalar {\bf
  5-plet} with $Y=1$. Note that, for some of the fields, particular
symbols are common in the literature, such as $\nu_R$, $\Delta$ and
$\Sigma$ for the type-I, type-II and type-III seesaw.

\subsection{Tree-level $d=5$}
\label{subsect:tree5}

As noted in \cite{Ma:1998dn}, there are only three possibilities to
de-construct the Weinberg operator at tree-level. A seesaw type-I is
generated via the introduction of a right-handed neutrino,
$\nu_R\equiv {\bf 1}^F_0$
\cite{Minkowski:1977sc,Yanagida:1979as,Mohapatra:1979ia}. It generates
Dirac mass terms for the active neutrinos via ${\bar \nu_R}H L$.  For
the singlet a Majorana mass term is allowed, $M_M{\bar \nu_R^c}\nu_R$,
which implies $\Delta L=2$.  The type-II seesaw requires a scalar
triplet $\Delta \equiv {\bf 3}^S_1$
\cite{Schechter:1980gr,Mohapatra:1980yp,Cheng:1980qt}. Here, the
simultaneous presence of the couplings $L \Delta L$ and $H
\Delta^{\dagger} H$ violates lepton number $\Delta L=2$. And, finally,
a type-III seesaw \cite{Foot:1988aq} can be generated with a fermionic
triplet $\Sigma \equiv {\bf 3}^F_0$. Here, the vector-like mass term
$M_{\Sigma} {\bar \Sigma^c}\Sigma$ is the source of the lepton number
violation.  Note, that type-I and type-III seesaw are generated
by the same topology, reducing the total number of $d=5$ topologies at
tree-level to two.

\subsection{Tree-level $d=7$}
\label{subsect:tree7}

At $d=7$ level one can construct five different topologies, one of
which however can not lead to any renormalizable model.  The
remaining four topologies have been discussed in \cite{Bonnet:2009ej}.
Only one of these topologies can generate a genuine $d=7$ neutrino mass
model in our sense, see figure (\ref{fig:treed7}). All other models
will require additional symmetries to avoid $d=5$ tree-level
neutrino masses.

\begin{center}
\begin{figure}[tbph]
\begin{centering}
\includegraphics[scale=1.0]{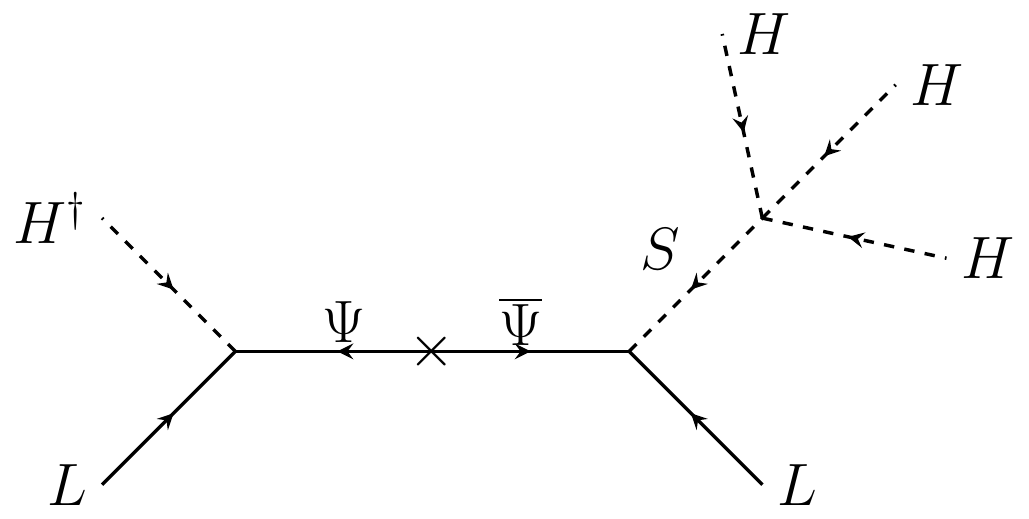}
\end{centering}
\protect\caption{\label{fig:treed7} Tree-level $d=7$ neutrino 
mass diagram, for the model described in \cite{Babu:2009aq}. 
$\Psi={\bf 3}^F_1$ and $S={\bf 4}^S_{3/2} $.}
\end{figure}
\par
\end{center}

This genuine $d=7$ tree-level model, BNT model in the following, was
first discussed in \cite{Babu:2009aq}. For a discussion of lepton
flavour violation in the BNT model see \cite{Liao:2010rx}.  The model
requires two new particles beyond the SM field content: (i) A
(vector-like) triplet fermion, $\Psi={\bf 3}^F_1$. \footnote{Its vector partner
${\bf 3}^F_{-1}$ is needed for a (effectively lepton number violating)
mass term.} And (ii) a scalar quadruplet $S \equiv {\bf 4}^S_{3/2}$. Note
that the quadruplet is the smallest representation which allows a
contraction $S^0 (H^0)^3$.

\subsection{1-loop $d=5$}
\label{subsect:1lp5}

In addition to the tree-level, there are many 1-loop $d=5$ models.
The classical example is the Zee model \cite{Zee:1980ai}. A systematic
analysis of all 1-loop $d=5$ topologies has been given in
\cite{Bonnet:2012kz}. In total, 6 topologies where found, but only two
of them (denoted as T-1 and T-3) can give genuine models in our
sense. All other topologies lead either to non-renormalizable models
or diagrams with infinite loop integrals (thus representing loop
corrections to tree-level quantities) \footnote{Topology T-4 has two
  divergent and two finite diagrams.  In \cite{Fraser:2015mhb} the
  diagram T-4-2-ii was used to generate a coupling $L \Delta L$ at
  one-loop. This diagram is classified as divergent in
  \cite{Bonnet:2012kz}. However, in \cite{Fraser:2015mhb} {\em two}
  diagrams of this type appear, with the infinity cancelled between
  diagrams. This can not be justified in terms of symmetry, instead it
  is due to the fact that lepton number is broken {\em softly} in the
  model of \cite{Fraser:2015mhb}.}  or can be understood as finite 
1-loop realizations of some particular vertex of one of the tree-level
$d=5$ seesaws. Topologies T-1 and T-3 lead to a total of four diagrams
shown in fig. (\ref{fig:loopd5}).  The Zee model \cite{Zee:1980ai}
falls within category T-1-ii, the scotogenic model of Ma
\cite{Ma:2006km} is an example of T-3.

\begin{center}
\begin{figure}[tbph]
\begin{centering}
\includegraphics[scale=0.6]{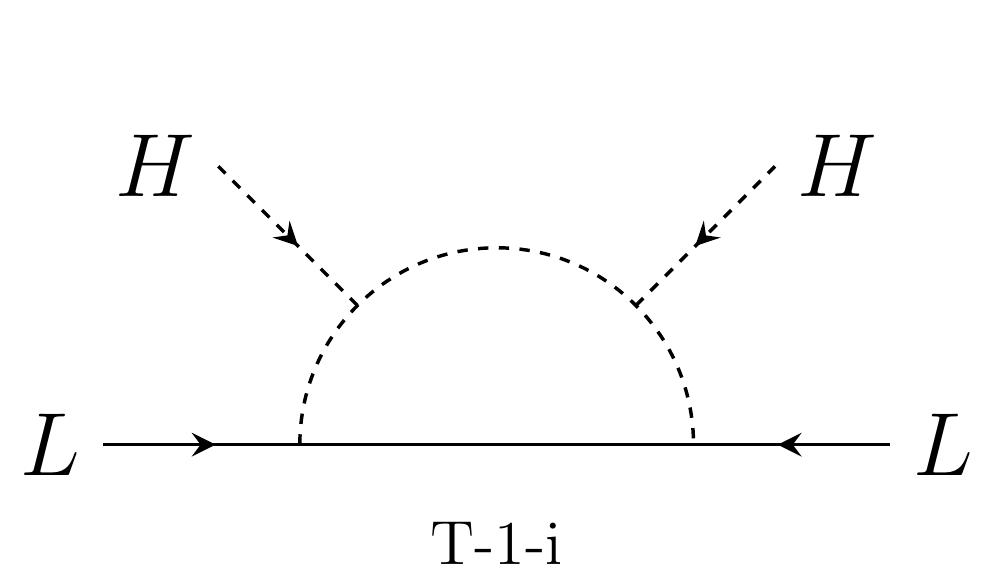}
\includegraphics[scale=0.6]{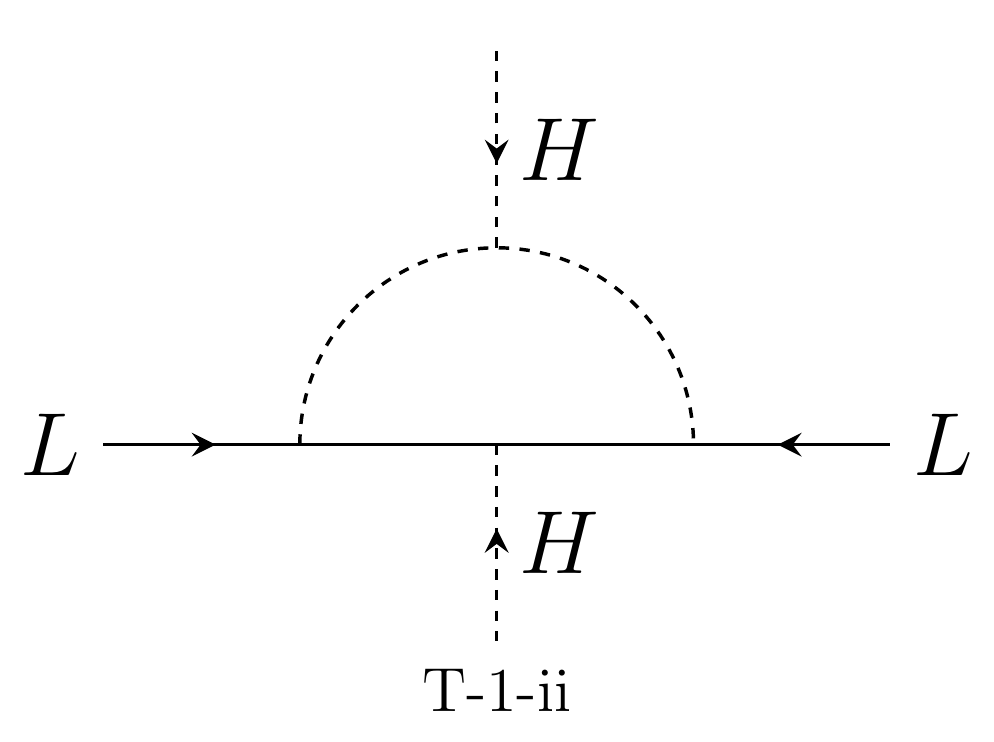}
\vskip3mm
\includegraphics[scale=0.6]{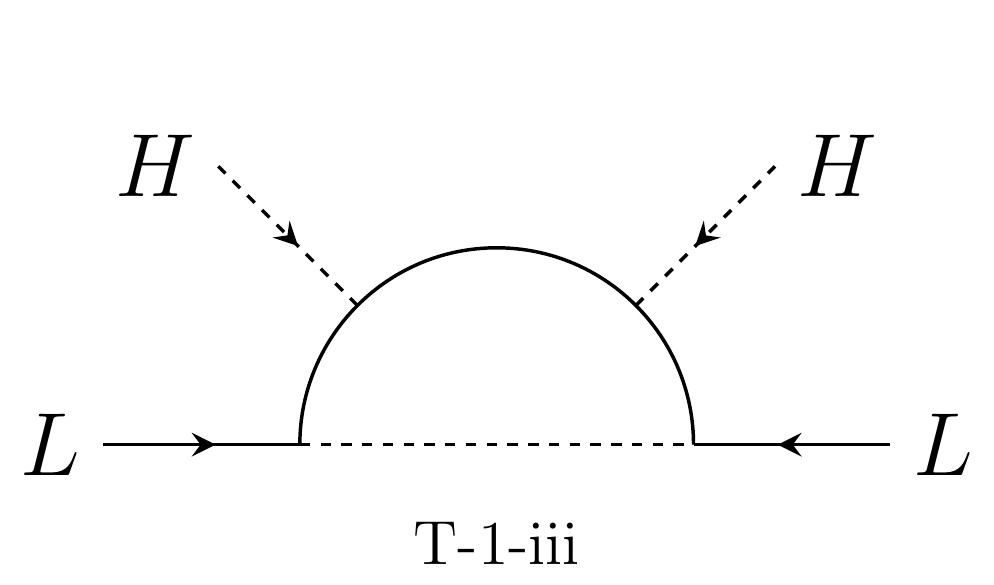}
\includegraphics[scale=0.6]{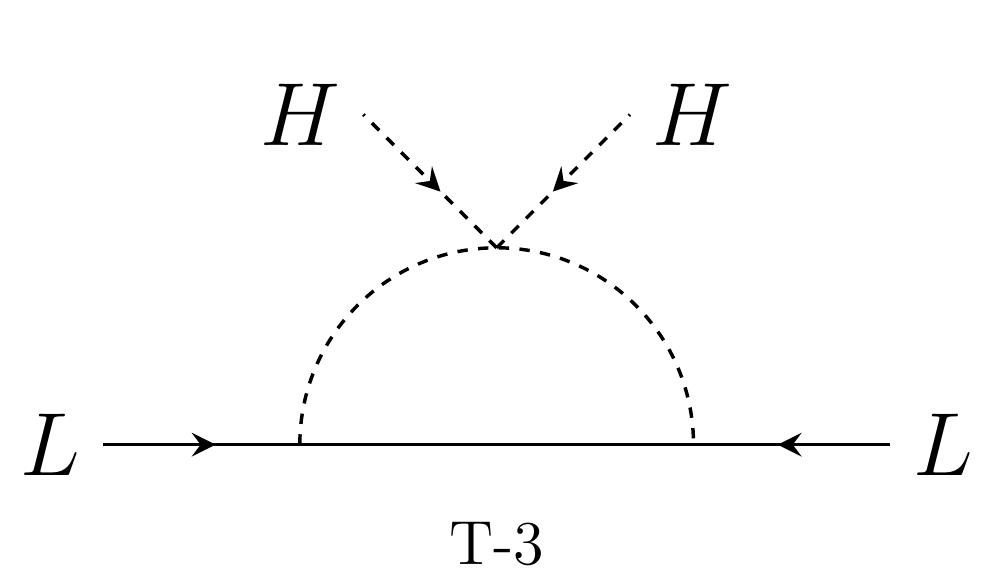}
\end{centering}
\protect\caption{\label{fig:loopd5} The four genuine $d=5$
  1-loop neutrino mass diagrams in the notation of \cite{Bonnet:2012kz}.}
\end{figure}
\par
\end{center}

While at tree-level the size of the representations as well as the
hypercharge of the new fields is fixed, at loop level there always
exists a ``tower'' of possible models. This is easily
understood. Consider, for example, the diagram T-3. The outside
leptons couple to a scalar and a fermion. Since $L$ is a $SU(2)$
doublet, the representation of the scalar and the fermion can be:
${\bf 1}\otimes {\bf 2}$, ${\bf 2}\otimes{\bf 3}$, ${\bf 3}\otimes
{\bf 4}$, etc. Similarly at the four scalar vertex the smallest
possibility is ${\bf 2}\otimes {\bf 2}$, but larger representations
can be inserted, with the only constraint that the product of the two
scalars can build a triplet. In the same way, the hypercharge of the
internal particles is fixed only up to an additive constant $x$, that
runs in the loop.\footnote{Of course, not all choices of $x$ will lead
  to phenomenologically acceptable models.}

The minimal possibility to build a model for T-3 is then that the
fermion is a $\nu_R={\bf 1}^F_0$, while the scalar is a doublet
${\bf 2}^S_{1/2}$, i.e. the well-known scotogenic model. This is minimal
in the sense that it uses the smallest representations and the
smallest value of the hypercharge possible, i.e. $x=0$. However,
${\bf 2}^S_{1/2}$ can not be the SM Higgs, it must be an additional
inert doublet.

Note that in our sense, strictly speaking, the scotogenic model is not
a ``genuine'' model, since it requires an additional symmetry (in the
minimal case a simple $Z_2$) to avoid the tree-level $d=5$ type-I
seesaw. This does not mean, however, that topology T-3 is non-genuine:
This topology has a finite loop integral and thus, no tree-level
counter term is needed in models that generate this topology.  Rather,
in order to avoid the type-I seesaw contribution without the use of an
additional symmetry, requires us to introduce fields with larger
hypercharges. The smallest possible choice is: ${\bf 1}^F_{1}$
(together with its vector partner) and two scalars ${\bf 2}^S_{1/2}$ and
${\bf 2}^S_{3/2}$.

\section{Classification\label{Sect:Class}}

In this section, we will discuss the classification of the different
$d=7$ 1-loop diagrams. We first construct all possible 1-loop
topologies with six external legs and then discard in different steps
those topologies that can not lead to genuine models. For the
remaining topologies we order all possible diagrams into different
classes, depending on the minimum size of the largest required
$SU(2)_L$ representation appearing in the corresponding diagram.

We note in passing that we will not discuss colour in detail, because
colour assignments can be trivially added: All particles outside loops
must be necessarily colour singlets, while pairs of particles in loops
can always be assigned colour in combinations ${\bf X}+{\bf\bar X}$,
for ${\bf X}={\bf 1}$, ${\bf 3}$, $\cdots$, which then couple to
``outside'' colour singlet particles.

\subsection{Topologies}
\label{subsect:topo}

We construct all possible 1-loop topologies with six external legs,
discarding from the start all self-energy corrections. This
construction can be done in different ways and we used two different
procedures to assure that all possible topologies were found. In total
there are 48 possible topologies. The complete list is shown in the
appendix.

We will briefly describe the methods we used to find the topologies.
The first procedure consists in taking the five $d=7$ tree-level
topologies and to generate loops in all possible combinations,
connecting either lines to lines or lines to vertices or vertices to
vertices, using only 3-point and 4-point vertices. From this list one
has to discard in the end all duplicates.

\begin{figure}
\begin{center} 
\includegraphics{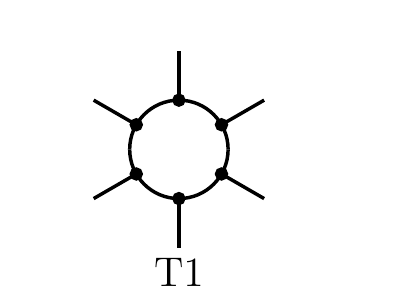}
\includegraphics{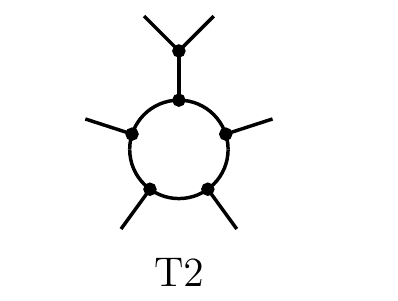}
\includegraphics{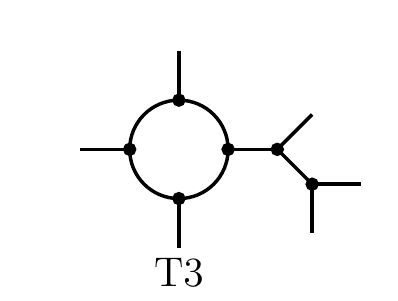}
\end{center}
\caption{Generating topologies, starting from the simplest topology,
  containing only 3-point vertices with all 6 particles connected to
  the loop, T1. Subsequent topologies are found by removing systematically 
  particles attached to the loop and reconnecting them to ouside
  particles, as shown for the examples T2 and T3. See text.}
\label{fig:T1T2T3}
\end{figure}

\begin{figure}
\centering 
\includegraphics{./figures/T1.pdf}
\includegraphics{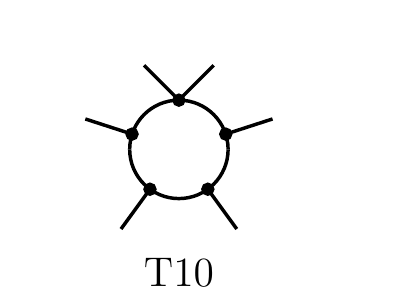}
\caption{Constructing topologies with 4-point vertices from existing
  topologies with only 3-point vertices by ``shrinking'' one
  connecting line. Here shown for the example how T1 generates T10.
  See also text.}
\label{fig:T1T10}
\end{figure}

The second procedure starts from the most simple realization of a
1-loop topology adding six external lines to the loop using only
3-point vertices, as shown in fig. (\ref{fig:T1T2T3}). From this
topology, all other toplogies can be found by systematically removing
lines attached to the loop and adding them to an outside particle
generating a new 3-point vertex, as shown in the figure for the
examples of T2, T3 etc. Once all possible topologies with only 3-point
vertices are found, all remaining topologies can be generated from the
earlier ones by shrinking one line connecting two 3-point vertices to
one new 4-point vertex, see the example fig. (\ref{fig:T1T10}).
Again, this procedure produces duplicates, which have to be identified
and discarded.

This second procedure provides a systematic construction of all
possible topologies in a more intuitive way than the first one
described above. It allows to classify topologies according to the
number of lines entering the loop, creating subgroups with the same
number of 4-leg vertices. This procedure is the one we use for
ordering the complete list of topologies, given in the appendix.

We then proceed to order topologies into different groups.  We can
discard immediately the six topologies shown in
\FIG{fig:TopologiesNRO}, because none of them can lead to a
renormalizable model. The next step is to generate all possible
diagrams and check if any field which generates neutrino masses at
lower order is required.

\begin{figure}
\centering 
\includegraphics{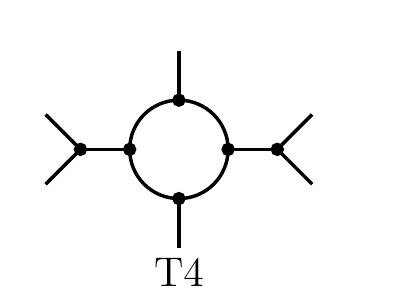}
\includegraphics{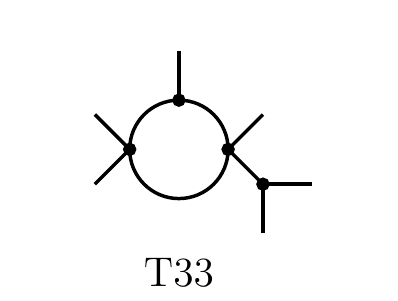}
\caption{Two example $d=7$ topologies which always will be accompanied
  by a tree-level seesaw $d=5$ contributions to the neutrino mass
  matrix.  For discussion see text.}
\label{fig:T4T33}
\end{figure}

Two examples of topologies, which always necessarily will be
accompanied by a tree-level $d=5$ seesaw, are shown in
\FIG{fig:T4T33}. These diagrams can be easily understood.  Every
topology with at least two 3-leg vertices on two external lines will
always generate a vertex $LH{\bar \nu_R}$ (or $LH{\bar \Sigma}$ or
$H\Delta^\dagger H$) and, thus a seesaw at tree-level, as in the
example $T4$ \FIG{fig:T4T33} on the left.  Topologies which contain
one 3-leg vertex with two external lines isolated by a 4-leg vertex
will always have a coupling of the type $L\Delta L$, as for example
the topology T33 in \FIG{fig:T4T33}, to the right. The 27 topologies,
for which all diagrams can be excluded due to this argument, are given
in \FIG{fig:TopologiesSeesaw}. The topologies T7, T22, T23 and T24 
in this figure are somewhat particular examples.  For these a
$\Delta$ always has to exist. One might think to bypass the $\Delta$
and introduce a quintuplet instead. However, the coupling of two
doublets to a 5-plet is zero due to $SU(2)_L$.

Before moving on a brief comment might be in order. Integrals for the
diagrams in topologies T4 and T33 are finite. One might therefore
wonder, whether it is possible to forbid one of the ``ingredients'' of
the tree-level $d=5$ seesaw, say one particular vertex, via a discrete
symmetry, only to generate it at 1-loop order. This was discussed at
length for $d=5$ 1-loop diagrams in \cite{Bonnet:2012kz}. At the $d=7$
level, however, this will not be possible, since $H^{\dagger} H$ is a
singlet under any discrete symmetry.

\begin{figure}
	\centering
	\includegraphics[scale=0.6]{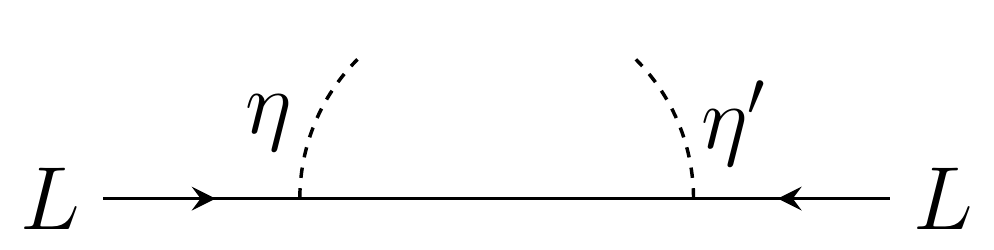}
	\caption{The particular piece of diagram that generates the
          1-loop $d=5$ diagram T-3, see \FIG{fig:loopd5}. If this
          structure exists in any diagram, the vertex with the two
          scalars $\eta$, $\eta'$ and two Higgses also always
          exist. This structure appears in many diagrams of the $d=7$
          topologies.}
	\label{fig:piece_dim5-1l}
\end{figure}

Next we turn to identifying topologies which generate diagrams
reducible to 1-loop $d=5$ models. This is not as straightforward as
the tree-level case. In particular, in this class of topologies many
diagrams lead to $d=5$ tree-level models, while only the remaining
diagrams can lead to $d=5$ 1-loop models. However, when the topology
is highly symmetric, as for example in $T1$, one can always find a
coupling between two internal fields and an external field which
bypasses the $H^\dagger$, giving one of the diagrams of
\FIG{fig:loopd5}. In addition, any diagram containing the structure
given in \FIG{fig:piece_dim5-1l} can be reduced to the well-known
diagram T-3 in \FIG{fig:loopd5}.  We list all topologies excluded due
to these arguments in fig. \ref{fig:Topos_d=5_1-loop}.

Then, there are diagrams which always contain the fields
$\textbf{4}^S_{3/2}$ and $\textbf{3}^F_1$, responsible of generating
neutrino mass at tree-level $d=7$ \cite{Babu:2009aq}. Examples are
diagrams of the topologies $T25$, $T29$ and $T35$
(\FIG{fig:Topologiesd=7tree}), which are excluded as genuine ones due
to this reason.

\begin{figure}
	\centering 
	\includegraphics[scale=1.2]{./figures/T10.pdf}\hskip15mm
	\includegraphics[scale=0.4]{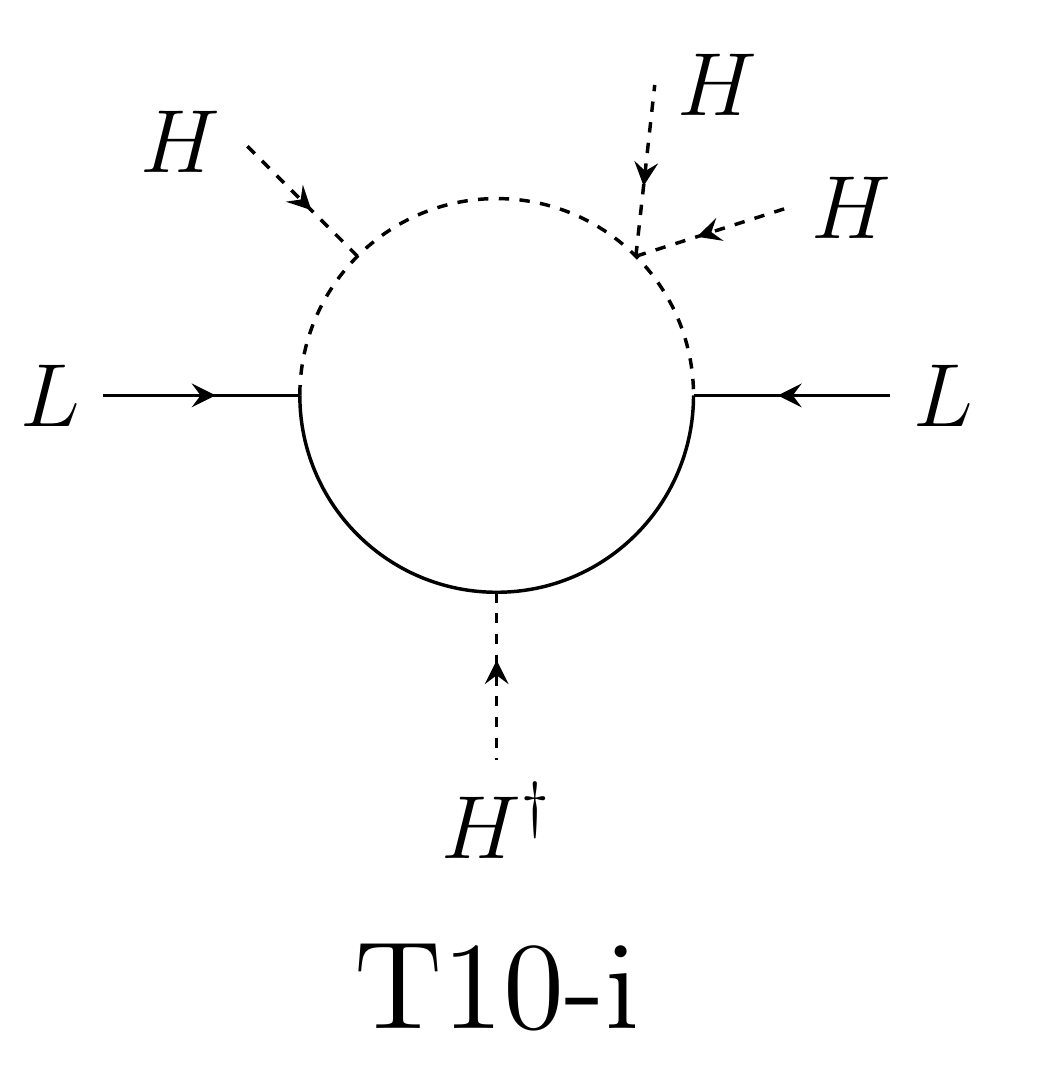}
	\caption{Topology T10 (left) and the only diagram derived from
          this topology, that can give a genuine model (right).}
	\label{fig:T10}
\end{figure}

Finally, in the remaining 8 topologies that are not completely
excluded by one of the above arguments, many but not all the diagrams
do not lead to genuine models.  For instance, from the 10 different
diagrams that one can generate from topology $T10$ (\FIG{fig:T10}
left), only one is not reducible to 1-loop $d=5$ (\FIG{fig:T10}
right).

In summary, from the initial 48 topologies only 8 have at least one
genuine 1-loop $d=7$ diagram. The excluded topologies are listed in
the appendix in figs \ref{fig:TopologiesSeesaw} - \ref{fig:TopologiesNRO}.
The next step is to classify the surviving topologies in terms of the
minimal $SU(2)_L$ representation needed to realize a genuine model.

\subsection{Diagrams: Minimal $SU(2)_L$ representations}

In all $d=7$ diagrams, in order to avoid neutrino masses at lower
order, a minimal size for the $SU(2)_L$ representations of the model
is required. We order the possible models according to the largest
representation present in a given model. The ``smallest'' or minimal
model that one can construct is then a model in which no
representation larger than $SU(2)_L$ triplets is needed. The next
smallest possibility is models with quadruplets. Here, one can
distinguish three different subgroups: (i) diagrams in which one
quadruplet is needed inside the loop; (ii) diagrams in which one
quadruplet appears outside the loop and internal particles need not be
larger than triplets; and (iii) models in which at least two
quadruplets are needed. We will discuss these three possibilities in
reverse order and then proceed to briefly discuss the triplet diagram.

For external fields, finding the minimal representation is
straightforward. A recurrent example in most of the diagrams of the
appendix is that of the vertex $HH^\dagger$-scalar. Since $2 \otimes 2
= 3 + 1$, the scalar could be a trivial singlet or the triplet $\phi
\equiv {\bf 3}^S_0$. The former case is directly reducible to a $d=5$
diagram, the latter is the one we are interested in.  The same
principle applies to the diagrams given in \FIG{fig:Diags4pletsInOut},
case (iii), for which the largest necessary representation is a
quadruplet. In order to avoid lower order contributions, one needs the
quadruplet $\textbf{4}^S_{1/2}$ ($\textbf{4}^F_{-1/2}$) outside the
loop. Moreover, one should be able to distinguish between these
quadruplets and a Higgs or a lepton doublet. For this reason, the
external quadruplet must couple to a singlet and another quadruplet
running inside the loop. All the diagrams of this type are depicted in
\FIG{fig:Diags4pletsInOut} and they always contain two quadruplets,
one outside the loop and another inside.

As we are dealing with the operator $LLHHHH^\dagger$, the maximum
hypercharges of an external quadruplet is $3/2$, i.e. $S$. These
diagrams corresponds to group (ii) defined above.  All the diagrams
given in \FIG{fig:Diags4pletsS} contains this scalar entering the loop
and they belong to the same topology $T16$ (\FIG{fig:TopoGenuine}).
Note that the hypercharge $3/2$ of $S$ prevents the possibility to
reduce these models to $d=5$ 1-loop.

\begin{figure}
	\centering
	\includegraphics[width=0.25\textwidth]{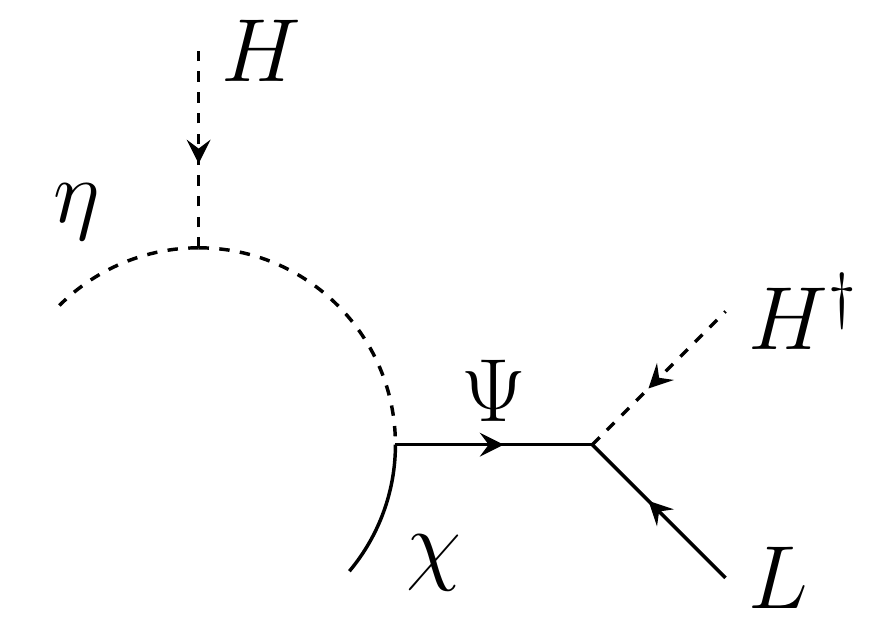}
	\qquad \qquad
	\includegraphics[width=0.17\textwidth]{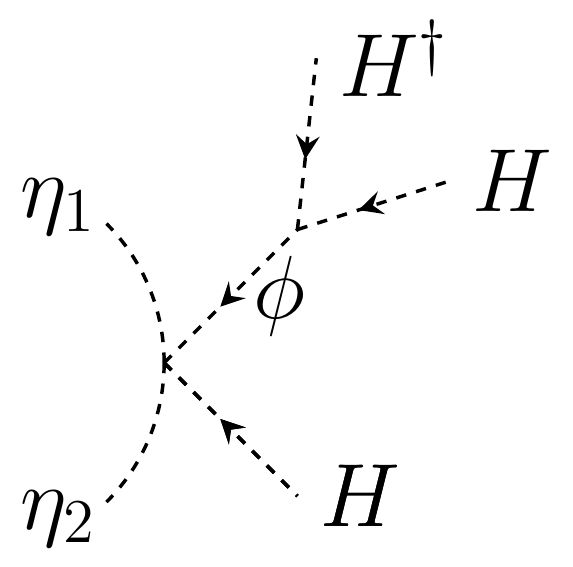}
	\caption{Structures that appear in topologies T12 (left) and
          T13 (right) which require at least a quadruplet and a
          singlet running inside the loop to avoid lower order
          contributions.  For discussion see text.}
	\label{fig:piece_4plet}
\end{figure}

The rest of the diagrams do not contain a external quadruplet. In the
minimal case, all the diagrams given in \FIG{fig:Diags4pletsIn} just
need one quadruplet running in the loop. Diagrams generated from the
topologies $T2$, $T12$ and $T13$ with a triplet entering the loop have
all similar structures to those given in \FIG{fig:piece_4plet}. The
minimum representations for the fields ($\chi$, $\eta$) or ($\eta_1$,
$\eta_2$) in \FIG{fig:piece_4plet}, are then a singlet and a
quadruplet, in oder to prevent a coupling of these fields with a
lepton doublet $L$ or the Higgs $H$, respectively.

The remaining diagram T10-i of \FIG{fig:Diags4pletsIn} is a rather
singular case of this group (i), with only one internal
quadruplet. Given the isolated $H^\dagger$ and the upper asymmetric
structure with three Higgses, the representation between the
Higgs vertices needs to be (at least) a quadruplet, otherwise
a $d=5$ 1-loop contribution is possible.

\begin{figure}[b]
\centering 
\includegraphics{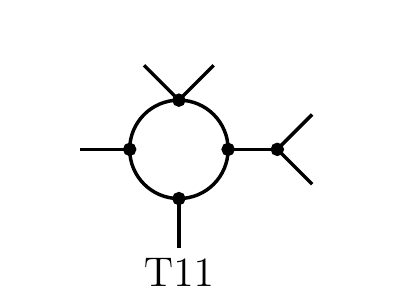}
\qquad \qquad
\includegraphics[scale=0.35]{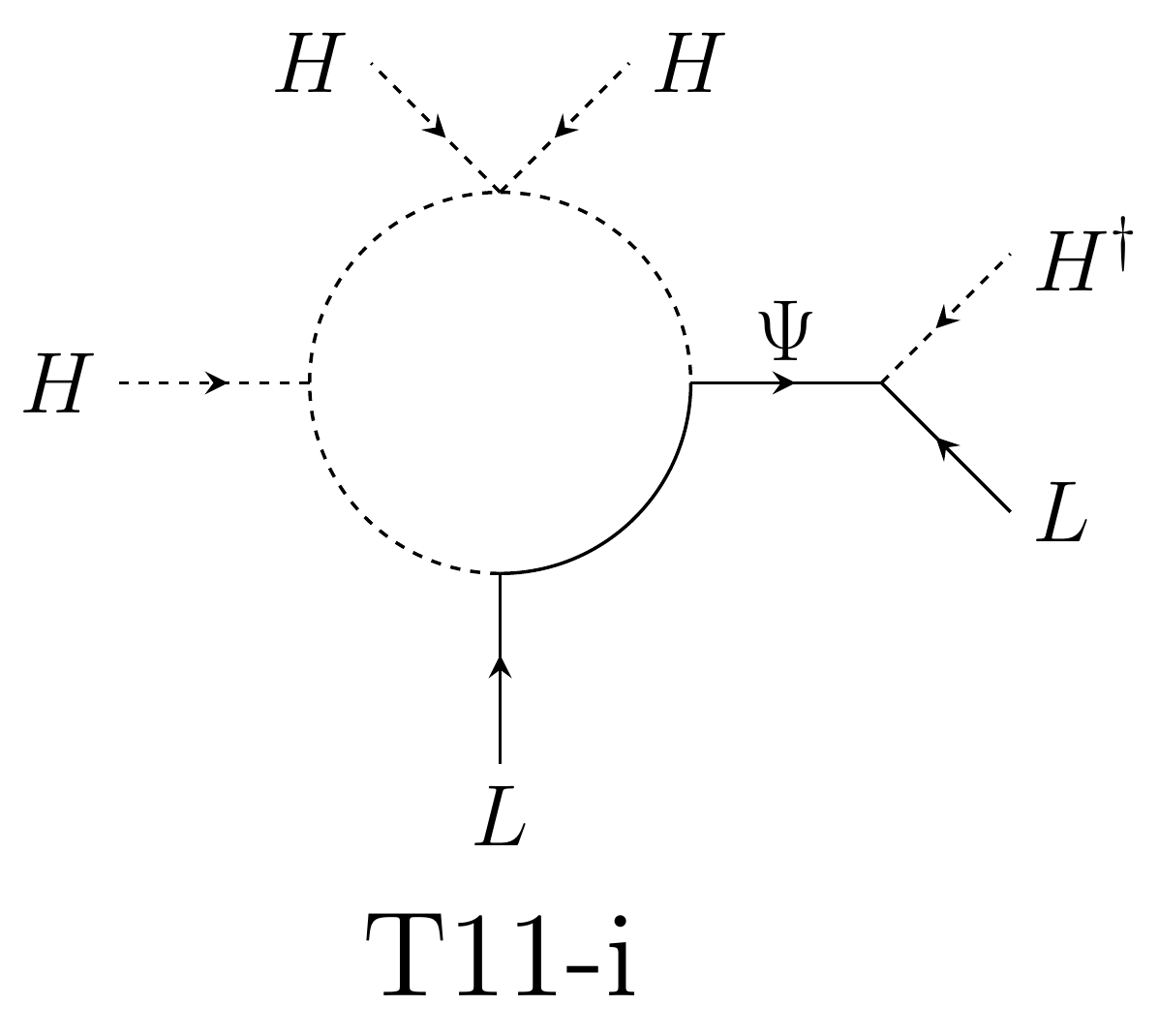}
\caption{Topology T11 to the left: The only topology for which
  a genuine model with no representations larger than triplets exist.
  The only genuine diagram for this topology is shown on the right.}
\label{fig:Triplet}
\end{figure}

After giving all the diagrams that can be construct with quadruplets
as the highest representation (figs
\ref{fig:Diags4pletsIn}-\ref{fig:Diags4pletsInOut}), the last case
depicted in \FIG{fig:Triplet} shows the only genuine diagram that can
be constructed with no representation bigger than triplet. Despite its
similarity to the structure given in \FIG{fig:piece_4plet} (left), the
4-leg vertex prevents lower order neutrino masses already with
triplets. The corresponding diagram \FIG{fig:Triplet} (right) with a
4-legs vertex followed by a triplet $\Psi$ cannot be bridged to
construct a 1-loop $d=5$ contribution given the relation between the
hypercharges of the fields running inside the loop.

To summarize, from the 8 genuine topologies, one can generate 23
diagrams. Among them, only one can be realized with no representation
larger than triplets as the maximum $SU(2)_L$ representation,
\FIG{fig:Triplet}. The other 22 diagrams generated from the 7
topologies given in \FIG{fig:TopoGenuine} can generate models with
representations up to quadruplets. This whole set can be divided
depending if the diagrams require one quadruplet running in the loop
(\FIG{fig:Diags4pletsIn}), outside the loop (\FIG{fig:Diags4pletsS})
or two quadruplets both inside and outside the loop
(\FIG{fig:Diags4pletsInOut}). Of course, models with larger
representations can be constructed and we will give one example in the
next subsection.

\subsection{Example models}
\label{subsect:models}

The complete list of diagrams, from which genuine $d=7$ 1-loop
models can be built is given in the appendix. Here, we will
briefly discuss three example models, which are among the most
simple models one can built from these diagrams. These models
are: (i) The simplest $d=7$ model, which requires no representation
larger than a triplet; (ii) one example model with an external 
quadruplet $S$; and (iii) an example model with an $SU(2)_L$
quintuplet. The latter serves to show, how models with larger
representations can easily be constructed from our list of
diagrams.

\subsubsection{Triplet model}

\begin{figure}
\centering
\includegraphics[scale=0.5]{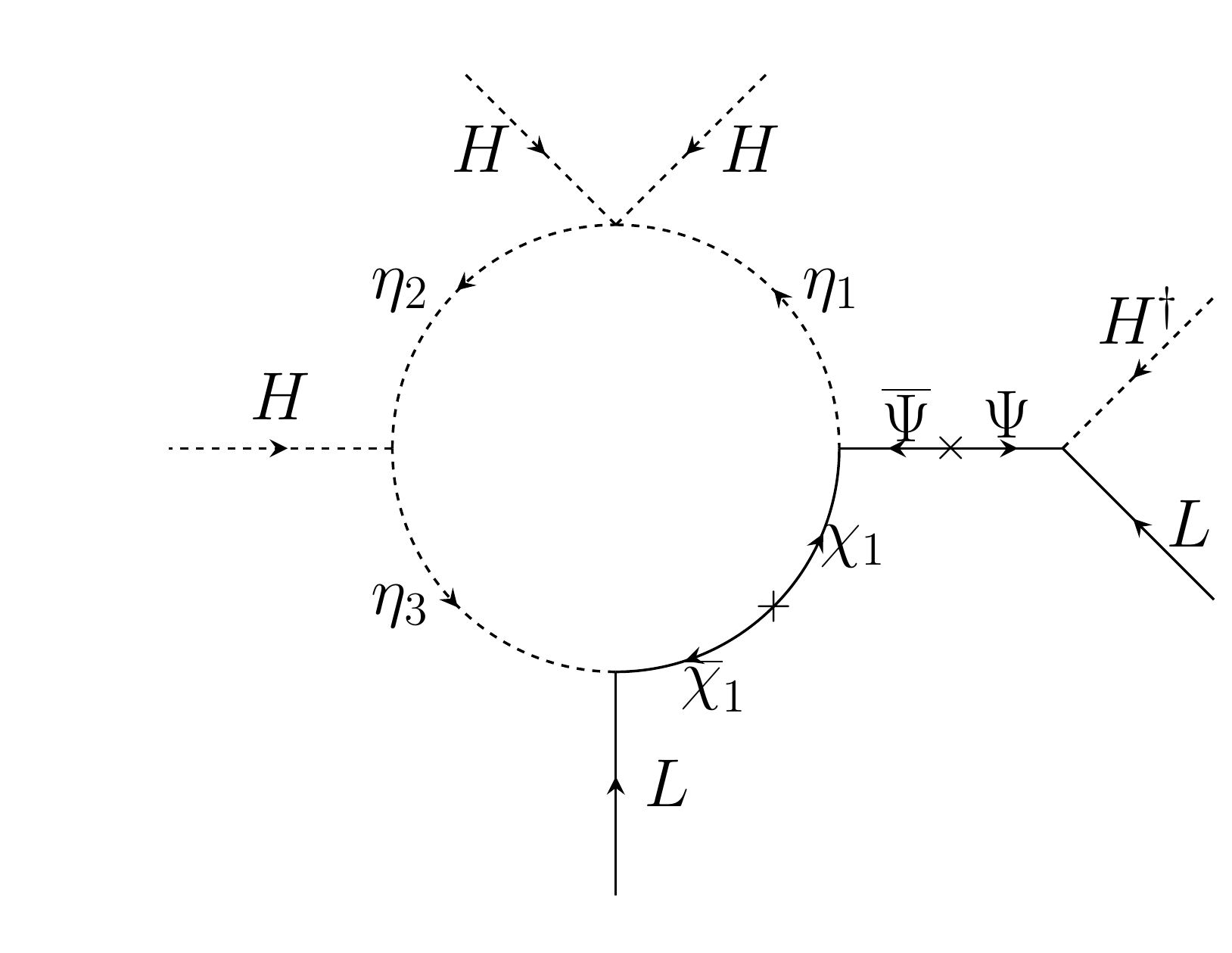}
\caption{The most minimal model that one can construct at 1-loop $d=7$
  order with no $SU(2)_L$ representations larger than triplet. The
  model is generated from the diagram T11-i in \FIG{fig:Triplet}.}
    \label{fig:3plet_model}
\end{figure}

As discussed above, there is only one possible diagram that has a
triplet as the largest $SU(2)_L$ representation, see \FIG{fig:Triplet}.
The model requires the fermionic triplet $\Psi=\textbf{3}^F_1$, that also
appears in the BNT model. A priori, for the particles inside the
loop hypercharge is not fixed. However, not all choices of hypercharge
will lead to genuine models, since lower order contributions
might appear. If we use only doublets and triplets inside the
loop, the smallest hypercharge assignments that lead to a
genuine model are:
\begin{center}
\begin{tabular}{ c c c }

$\Psi=\left(
\begin{matrix}
\Psi^{++} \\ 
\Psi^{+}  \\
\Psi^{0}
\end{matrix} 
\right) \sim \textbf{3}_1^F$

&\qquad

$\eta_1=\left(
\begin{matrix}
\eta_1^{++} \\ 
\eta_1^{+}
\end{matrix} 
\right) \sim \textbf{2}_{3/2}^S$

&\qquad

$\eta_2=\left(
\begin{matrix}
\eta_2^{+++} \\ 
\eta_2^{++}
\end{matrix} 
\right) \sim \textbf{2}_{5/2}^S$

\end{tabular}

\vspace*{0.5cm}

\begin{tabular}{ c c }

$\eta_3=\left(
\begin{matrix}
\eta_3^{++++} \\ 
\eta_3^{+++}  \\
\eta_3^{++}
\end{matrix} 
\right) \sim \textbf{3}_3^S$

&\qquad

$\chi_1=\left(
\begin{matrix}
\chi_1^{+++} \\ 
\chi_1^{++}
\end{matrix} 
\right) \sim \textbf{2}_{5/2}^F$.

\end{tabular}
\end{center}
The model generates neutrino masses via the diagram depicted in
\FIG{fig:3plet_model}. $\eta_1$ has the smallest hypercharge of the
particles in the loop. For colorless particles it is not possible to
find a smaller hypercharge assignment that leads to a genuine
model. For example, choosing $\eta_1 = {\bf 2}_{1/2}^S$ instead
would result in a model, which also has the diagram T-3 at $d=5$
level.

One interesting aspect of this model is that the scalar triplet
inside the loop has a component which carries 4 units of electric
charge.

\subsubsection{Quadruplet model}

While for triplets as the maximal representation there is only one
diagram, for quadruplets three distinct groups of model exist, as
discussed above. We choose an example based on an external quadruplet
$S={\bf 4}^S_{3/2}$. The example model we choose is based on diagram
T16-ii.

As in the triplet case, hypercharge and $SU(2)_L$ representation are
not uniquely fixed. The minimal model, again in the sense of using the 
smallest possible hypercharge assignment for colourless internal fields,
has the following particle content:
\begin{center}
\begin{tabular}{ c c c }

$S=\left(
\begin{matrix}
S^{+++} \\
S^{++} \\ 
S^{+}  \\
S^{0}
\end{matrix} 
\right) \sim \textbf{4}_{3/2}^S$

\quad & \quad

$\chi_1=\left(
\begin{matrix}
\chi_1^{++} \\ 
\chi_1^{+}
\end{matrix} 
\right) \sim \textbf{2}_{3/2}^F$

\quad & \quad

$\chi_2=\left(
\begin{matrix}
\chi_2^{++++} \\ 
\chi_2^{+++} \\ 
\chi_2^{++}
\end{matrix} 
\right) \sim \textbf{3}_3^F$

\end{tabular}
\begin{tabular}{ c c }

$\eta_1=\eta_1^{++} \sim \textbf{1}_2^S$.

\qquad & \qquad

$\eta_2=\left(
\begin{matrix}
\eta_2^{+++} \\ 
\eta_2^{++}
\end{matrix} 
\right) \sim \textbf{2}_{5/2}^S$.

\end{tabular}
\end{center}

\begin{figure}
\centering
\includegraphics[scale=0.5]{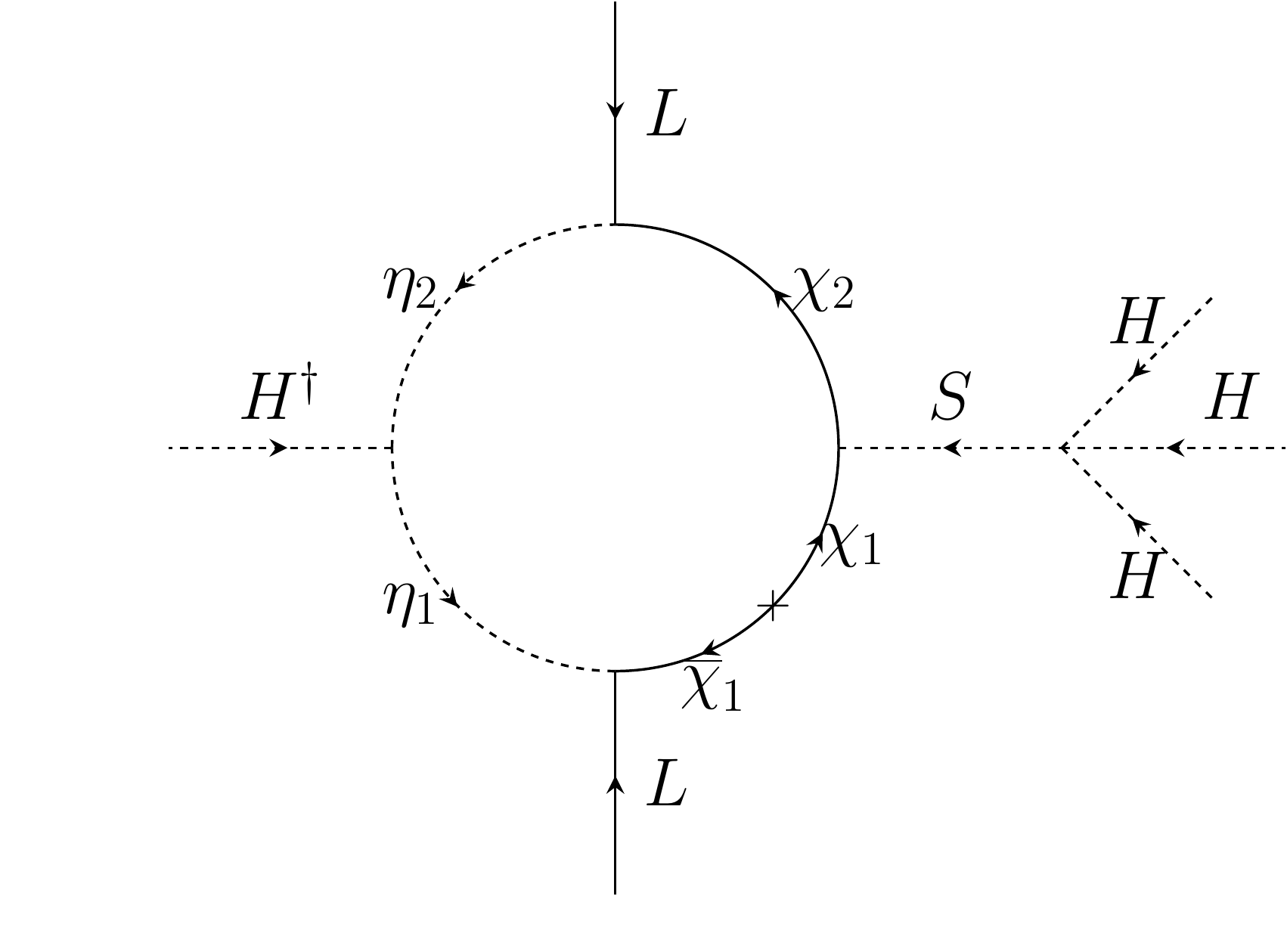}
\caption{Example of a $d=7$ 1-loop model with an external quadruplet
  $S={\bf 4}^S_{3/2}$, generated from the diagram T16-ii in
  \FIG{fig:Diags4pletsS}. This model contains only doublet and triplet
  representations inside the loop, see text.}
    \label{fig:4plet_model}
\end{figure}

This model generates neutrino masses via the diagram of
\FIG{fig:4plet_model}. In this example, lower order contributions can
be avoided due to the hypercharge of $S$. A model with smaller
hypercharges, for example $\eta_1$ chosen to be $\eta_1={\bf 1}_1^S$,
would again not be genuine, since it would necessarily have a $d=5$
1-loop contribution, i.e. the classical Zee model \cite{Zee:1980ai},
see diagram T-1-ii in fig.  \ref{fig:loopd5}. Note that, while the
triplet model contains a scalar with 4 units of electric charge, in
the quadruplet model it is an internal fermion that has such a large
electric charge.

\subsubsection{Quintuplet model}

\begin{figure}
    \centering
	\includegraphics[scale=0.5]{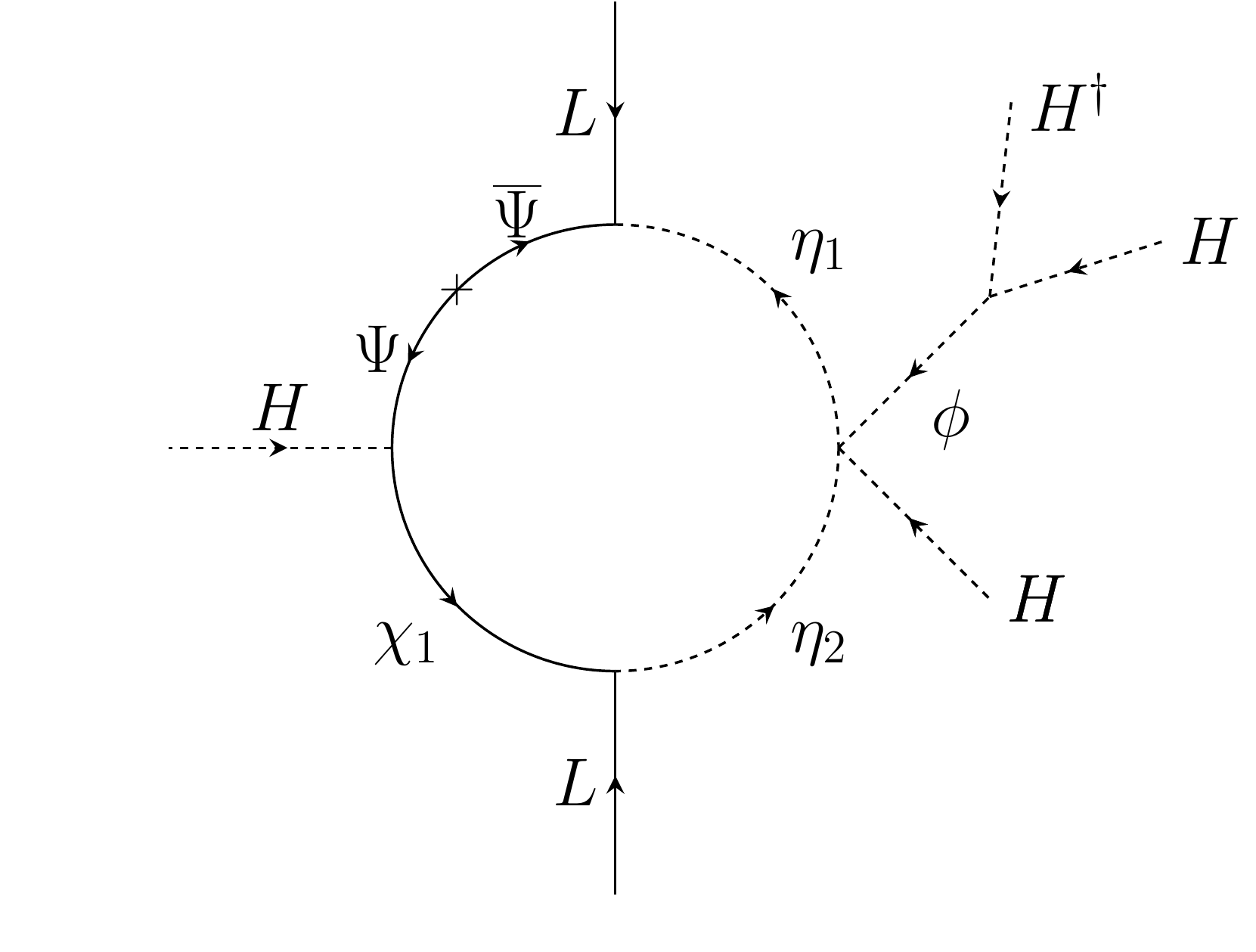}
    \caption{Example of one of the most minimal models that one can
      construct at $d=7$ 1-loop order with $SU(2)_L$ representations up
      to 5-plets generated from the diagram T13-i in
      \FIG{fig:Diags4pletsIn}.}
    \label{fig:5plet_model}
\end{figure}

Finally, our last example is a model based on diagram T13-i in
\FIG{fig:Diags4pletsIn}. It contains the field $\phi={\bf 3}_0^S$ and
a quintuplet in the loop. The diagram for the generation of the
neutrino masses is shown in \FIG{fig:5plet_model}. The minimal
particle content containing a 5-plet is given by:

\begin{center}
\begin{tabular}{ c c c }

$\phi=\left(
\begin{matrix}
\phi^{+}  	\\
\phi^{0}	\\
\phi^{-}
\end{matrix} 
\right) \sim \textbf{3}_0^F$

&\qquad

$\Psi=\left(
\begin{matrix}
\Psi^{++} \\ 
\Psi^{+}  \\
\Psi^{0}
\end{matrix} 
\right) \sim \textbf{3}_1^F$

&\qquad

$\chi_1=\left(
\begin{matrix}
\chi_1^{+++} \\ 
\chi_1^{++}  \\
\chi_1^{+}   \\
\chi_1^{0}
\end{matrix} 
\right) \sim \textbf{4}_{3/2}^F$
\end{tabular}

\vspace*{0.5cm}

\begin{tabular}{ c c }

$\eta_1=\left(
\begin{matrix}
\eta_1^{++}  \\ 
\eta_1^{+}  
\end{matrix} 
\right) \sim \textbf{2}_{3/2}^S$

&\qquad

$\eta_2=\left(
\begin{matrix}
\eta_2^{+++} \\ 
\eta_2^{++}  \\
\eta_2^{+}   \\
\eta_2^{0}   \\
\eta_2^{-}   
\end{matrix} 
\right) \sim \textbf{5}_1^S$.

\end{tabular}
\end{center}

The maximum representation in this model is a quintuplet, $\eta_2$.
Since it couples to the Higgs, $\phi$ and $\eta_1$, $\eta_1$ could be
either a ${\bf 2}$, ${\bf 4}$, ${\bf 6}$ or a ${\bf 8}$. However, only
for the case of $\eta_1$ being a doublet, a genuine model results.
This is because, once the coupling $\eta_1 H \eta_2^\dagger$ is
allowed, one can construct again a $d=5$ 1-loop diagram with the
particle content of the model.

It is worth noting that from three example models we have discussed,
the quintuplet model is the only one, in which the representations in
the loop contain a neutral component. For this model, one can thus
follow the idea of the scotogenic model  \cite{Ma:2006km}: Add a
discrete $Z_2$ symmetry to the model, under which the internal particles
are odd, and the lightest neutral particle can be a cold dark
matter candidate.

\section{Summary\label{Sect:Sum}}

We have discussed neutrino masses at 1-loop $d=7$ order. We have
identified all possible topologies that can lead to genuine models,
i.e. models that are not accompanied by either a $d=5$ or $d=7$
tree-level mass term nor by a $d=5$ 1-loop neutrino mass. We have
found that only 8 out of a total of 48 topologies can lead to genuine
models.

We then ordered the remaining, possibly genuine, diagrams into
different groups, depending on the minimal field content necessary to
construct a model. There is only one possible diagram for which the
largest necessary representation is a triplet. The remaining 7
topologies yield 22 diagrams, with the largest representation being at
least a quadruplet. We then briefly discussed three example models,
starting from the triplet model, with one additional example for a
quadruplet and one for a quintuplet each. 

To avoid lower order neutrino masses, the ``genuine'' models we
discussed always have to introduce five new multiplets, usually with
quite a large hypercharge for at least one of them. Thus, these $d=7$
models are necessarily more complicated constructions than the
classical seesaw. From a theoretical point of view this might make
these models less attractive.  However, in particular due to the large
electrical charges in these models, one can expect interesting
signatures for them at colliders. We reiterate that the $d=7$ 1-loop
contribution can only be dominant, if at least some of the new
particles have masses below roughly 2 TeV.

\bigskip
\centerline{\bf Acknowledgements}

\bigskip
We thank R. Fonseca for discussion. This work was supported by the
Spanish MICINN grants FPA2014-58183-P and Multidark CSD2009-00064
(MINECO), FPU15/03158 (MECD) and PROMETEOII/2014/084 (Generalitat
Valenciana). J.C.H. is supported by Chile grants Fondecyt No. 1161463,
Conicyt ACT 1406 and Basal FB0821.

\bigskip

\section{Appendix\label{Sect:App}}

In this appendix we present the list of all $d=7$ 1-loop topologies,
classified into genuine and non-genuine topologies, as discussed in
the main text. We also give the complete list of diagrams that can
lead to "genuine" $d=7$ neutrino mass models with $SU(2)_L$ quadruplet
representations.

\subsection{Topologies}
\label{subsect:app topo}

\begin{figure}[h]
\centering
\begin{tabular}{ c c c c }
\includegraphics{./figures/T2.pdf}  &  \includegraphics{./figures/T3.pdf}  &  
\includegraphics{./figures/T10.pdf}  &  \includegraphics{./figures/T11.pdf}
	\\  
\includegraphics{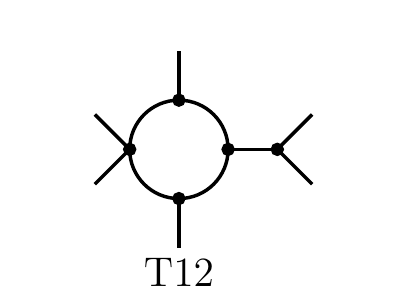}  &  \includegraphics{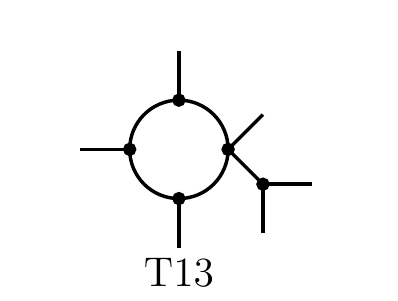}  
&  \includegraphics{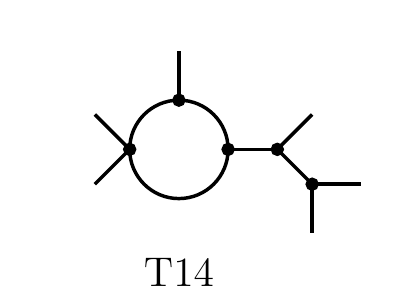} & \includegraphics{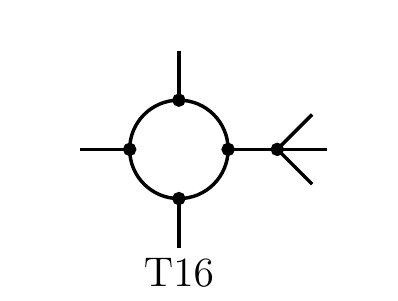}
\end{tabular}
\caption{Topologies that can lead to a genuine $d=7$ 1-loop neutrino
  mass model. T11 is the only topology for which the largest
  representation can be as small as a $SU(2)_L$ triplet. For all other
  topologies at least one quadruplet must appear in the diagram for
  the model to be genuine.  The quadruplet diagrams based on those
  topologies are shown in figs \ref{fig:Diags4pletsIn},
  \ref{fig:Diags4pletsS} and \ref{fig:Diags4pletsInOut}. For the
  triplet model see fig. \ref{fig:3plet_model}.}
 \label{fig:TopoGenuine}
\end{figure}

Fig. \ref{fig:TopoGenuine} shows the 8 topologies that {\em can} lead
to genuine $d=7$ 1-loop models. We stress again that not all diagrams,
derived from these topologies, are necessarily genuine, as discussed in
the main text. Note that only T11 can give a model in which the
largest representation can be as small as a $SU(2)_L$ triplet. All 
other 7 topologies require at least one quadruplet for genuine models.

\begin{figure}[h]
\centering
\begin{tabular}{ c c c c }
\includegraphics{./figures/T4.pdf}  &  \includegraphics{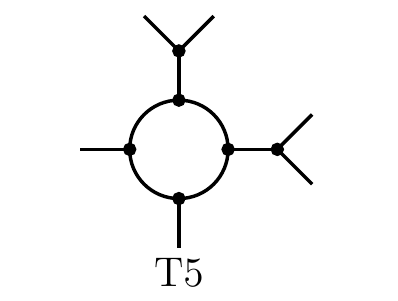}  &  
\includegraphics{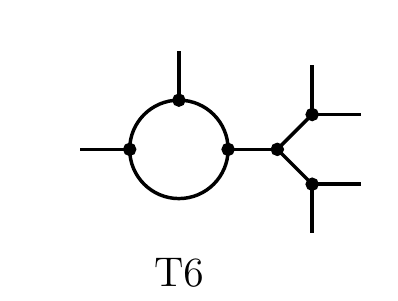}  &  \includegraphics{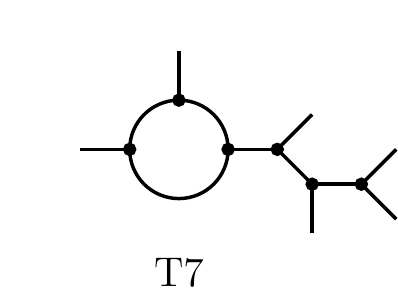}  
    		\\
\includegraphics{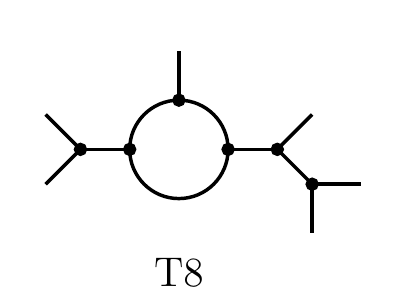}  &  \includegraphics{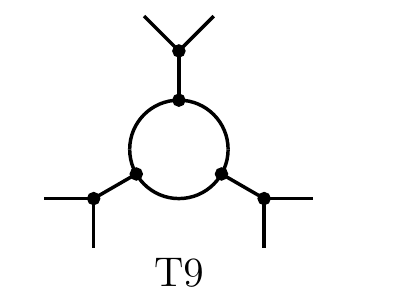}  &  
\includegraphics{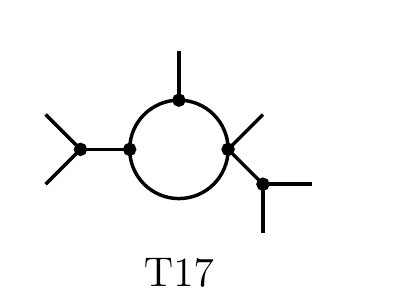}  &  \includegraphics{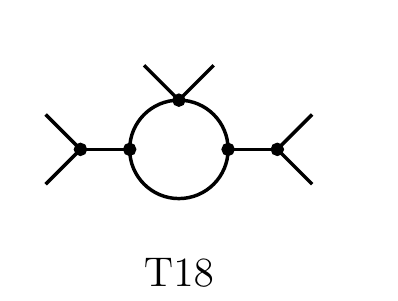}	
    		\\
\includegraphics{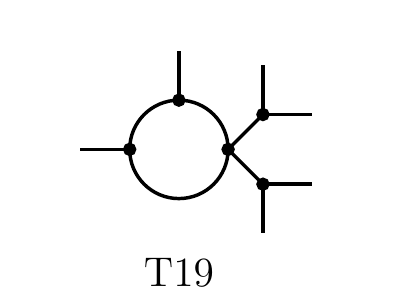}  &  \includegraphics{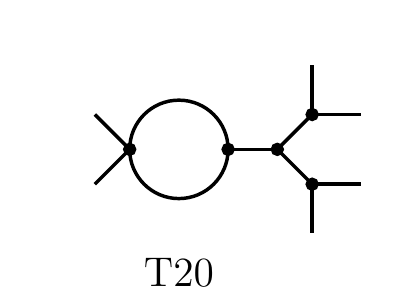}  &
\includegraphics{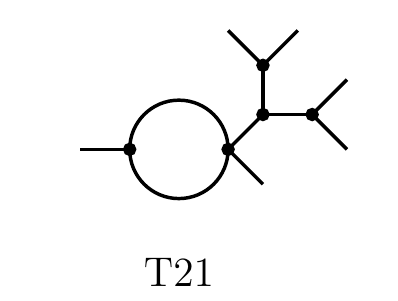} &  \includegraphics{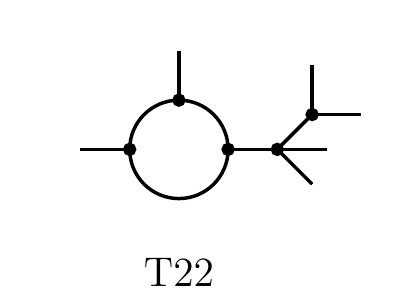}
    		\\
\includegraphics{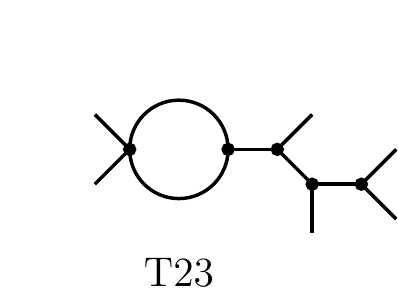}  &  \includegraphics{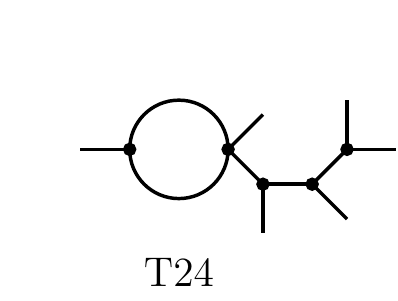} & 
\includegraphics{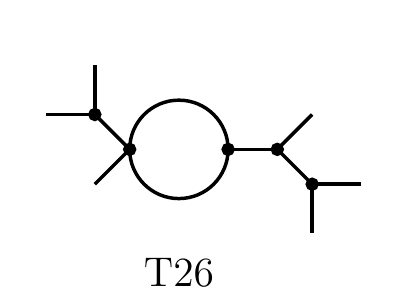}  &  \includegraphics{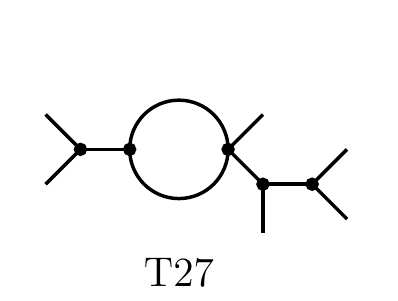}
    		\\
\includegraphics{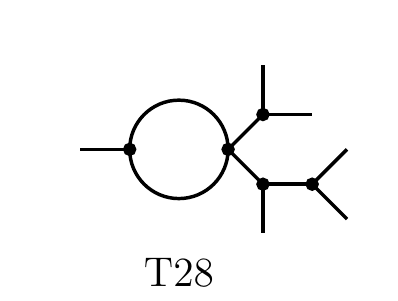}  &  \includegraphics{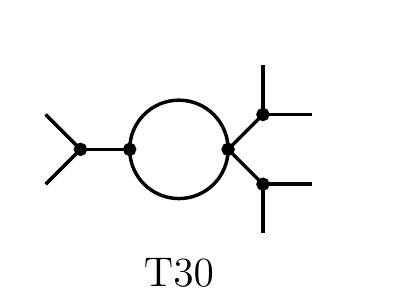} & 
\includegraphics{./figures/T33.pdf} & \includegraphics{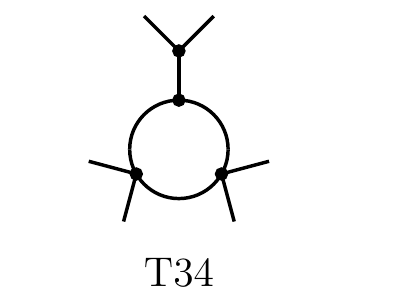}
    		\\
\includegraphics{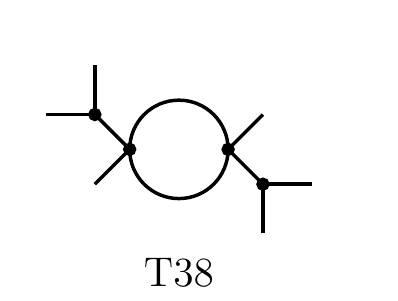}  &  \includegraphics{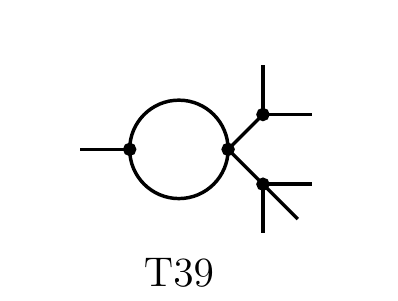} & 
\includegraphics{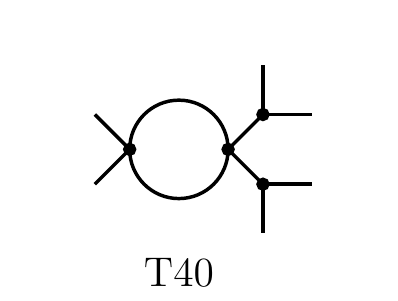}  &  \includegraphics{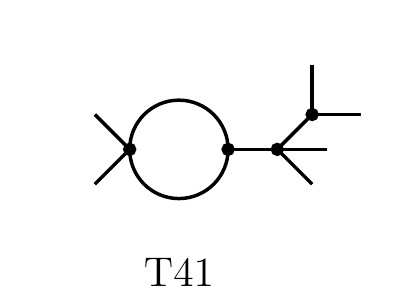}
    		\\
\includegraphics{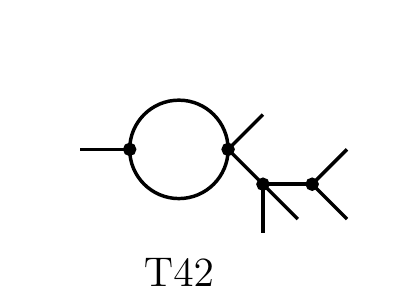}  &  \includegraphics{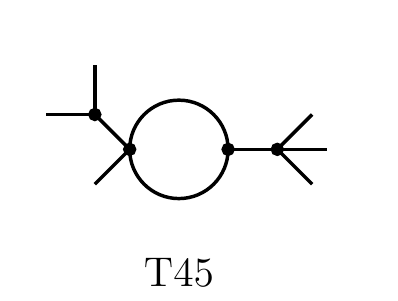} & 
\includegraphics{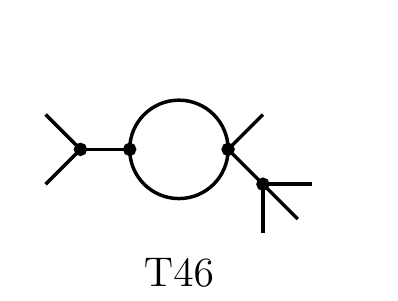}  &  
\end{tabular}
\caption{Topologies that necessarily lead to a $d=5$ tree level neutrino
  mass, see text. \label{fig:TopologiesSeesaw}}
\end{figure}

In fig. \ref{fig:TopologiesSeesaw} we list all topologies for which
{\em all} diagrams are excluded, since they contain either a singlet
fermion $\nu_R$ ($\textbf{1}_0^F$) or a triplet scalar $\Delta$
($\textbf{3}_{-1}^S$) or a triplet fermion $\Sigma$
($\textbf{3}_0^F$). All diagrams from these topologies thus will also
generate a tree-level $d=5$ seesaw contribution to the neutrino mass
matrix.

\begin{figure}[h]
\centering
\begin{tabular}{ c c }
   \includegraphics{./figures/T1.pdf}  &  \includegraphics{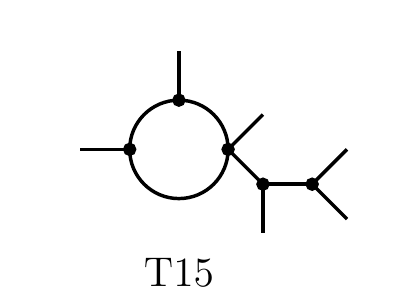}
   \\
   \includegraphics{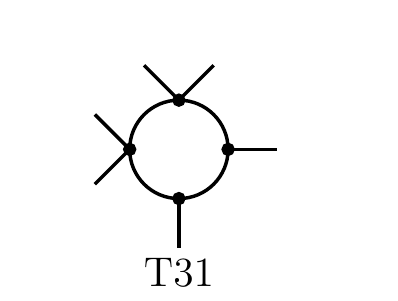}  &  \includegraphics{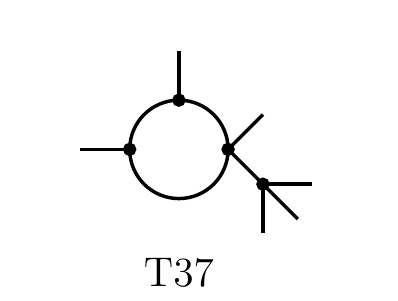}
\end{tabular}
\caption{Topologies that lead to a $d=5$ 1-loop neutrino mass. All
  diagrams generated from the topologies in this class, not already
  excluded because they generate a $d=5$ tree-level mass, include the
  particle content necessary to generate neutrino mass through one of
  the four genuine $d=5$ 1-loop diagrams, see fig. \ref{fig:loopd5}.
  Note that T15 is an exceptional case, since it has diagrams for all
  three possibilities: Tree-level $d=5$, 1-loop $d=5$ and tree-level
  $d=7$. }
  \label{fig:Topos_d=5_1-loop}
\end{figure}

In fig. \ref{fig:Topos_d=5_1-loop} we list the topologies for which 
many but not all diagrams are excluded by a $d=5$ tree-level seesaw. 
For these topologies {\em all remaining} diagrams are excluded because a 
1-loop $d=5$ contribution to the neutrino mass necessarily exists.

\begin{figure}[h]
\centering
\begin{tabular}{ c c c }
\includegraphics{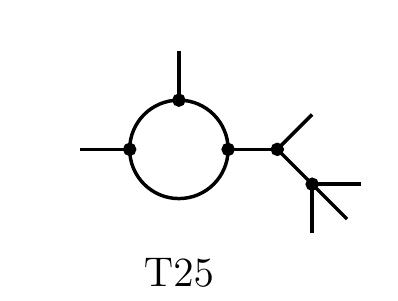}  &  \includegraphics{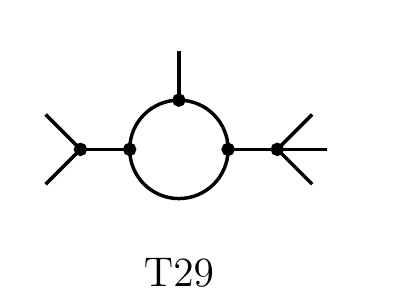}
& \includegraphics{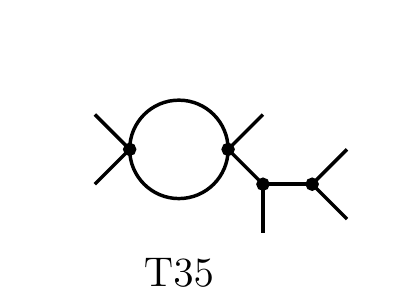} 
\end{tabular}
\caption{Topologies, which lead to a $d=7$ tree level neutrino
  mass. For each of these topologies one can construct diagrams, which
  have a $d=5$ tree-level mass. All remaining diagrams contain the
  scalar $S$ ($\textbf{4}^S_{3/2}$) along with the fermion $\Psi$
  ($\textbf{3}^F_1$) (\FIG{fig:treed7}) \cite{Babu:2009aq}. }
\label{fig:Topologiesd=7tree}
\end{figure}

In fig. \ref{fig:Topologiesd=7tree} we list topologies, which lead to
a $d=7$ tree level neutrino mass. For each of these topologies one can
construct diagrams, which have a $d=5$ tree-level mass. All remaining
diagrams, contain the scalar $S$ ($\textbf{4}^S_{3/2}$) along with the
fermion $\Psi$ ($\textbf{3}^F_1$) and thus generate the $d=7$
tree-level BNT model \cite{Babu:2009aq}.  We note in passing that one
can, in principle, use these diagrams to radiatively generate one of
the vertices in the BNT model. This is very similar to the discussion
for the radiative generation of a seesaw coupling given in
\cite{Bonnet:2012kz} at $d=5$ level.

\begin{figure}[h]
\centering
\begin{tabular}{ c c c c }
 \includegraphics{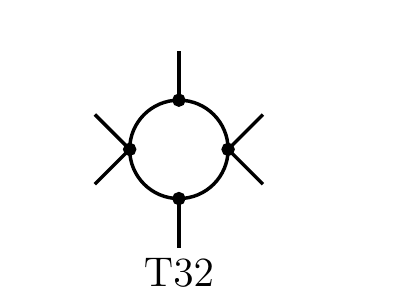}  &  \includegraphics{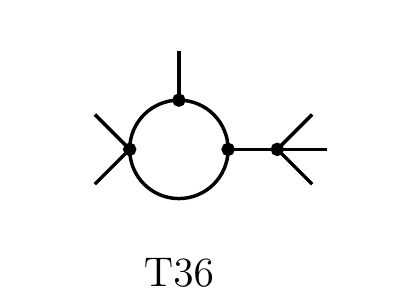} &
 \includegraphics{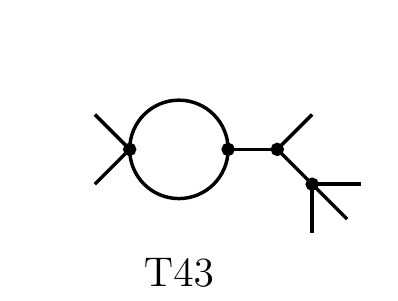}  &  \includegraphics{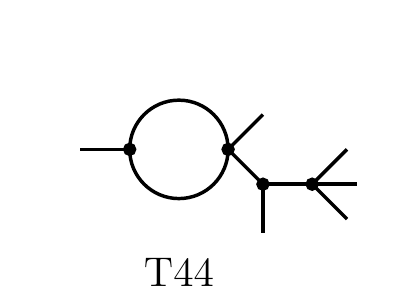} 
   \\
 \includegraphics{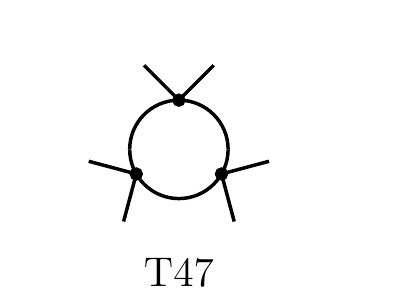}  &  \includegraphics{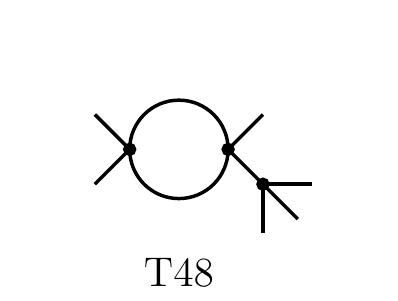}
\end{tabular}
\caption{Topologies discarded because they lead to non-renormalizable
  operators.}
\label{fig:TopologiesNRO}
\end{figure}

In fig. \ref{fig:TopologiesNRO} for completeness we show the
topologies which are excluded, since they can never lead to a
renormalizable model.

\subsection{Genuine diagrams}

\begin{figure}[h]
\centering
\begin{tabular}{ c c }
\includegraphics[scale=0.35]{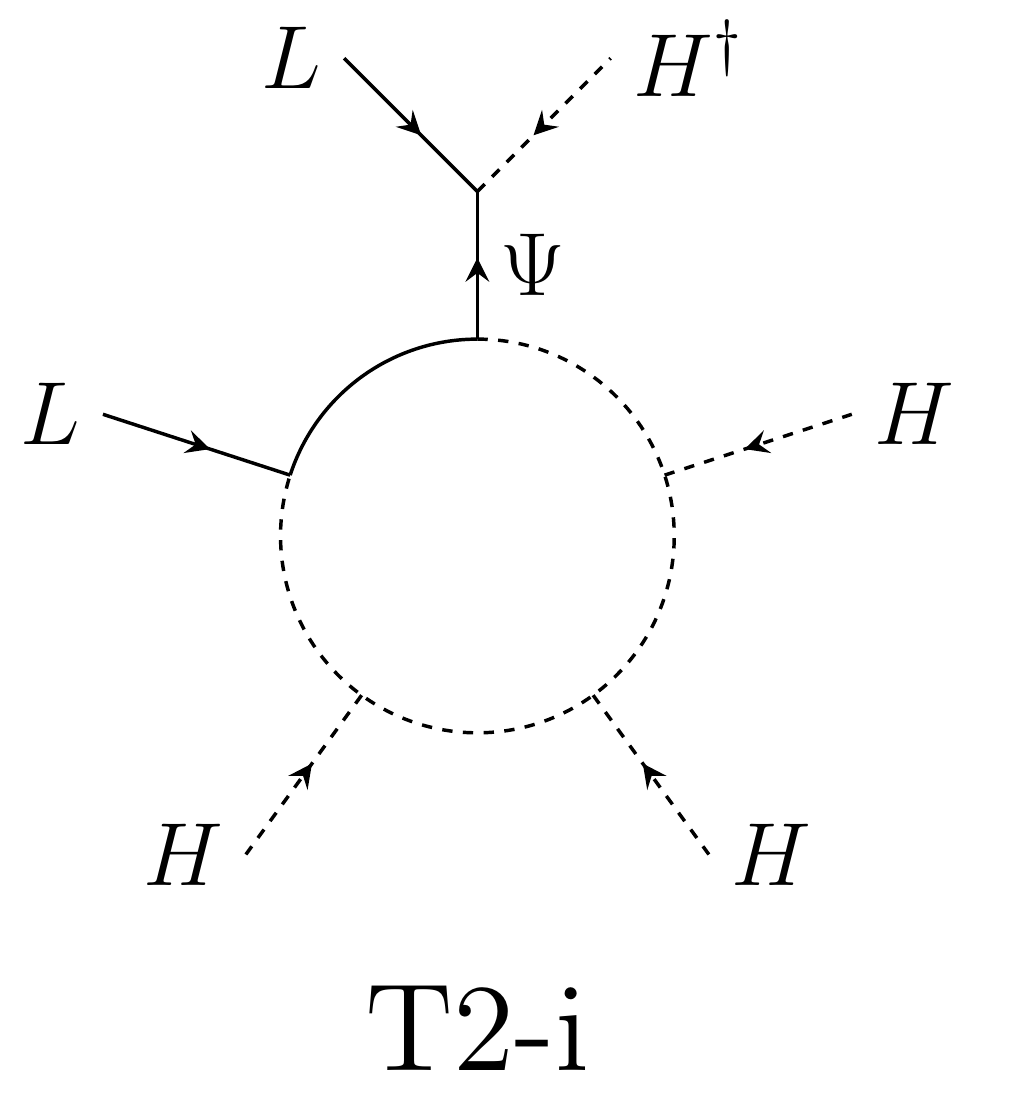} \quad &  \quad 
\includegraphics[scale=0.35]{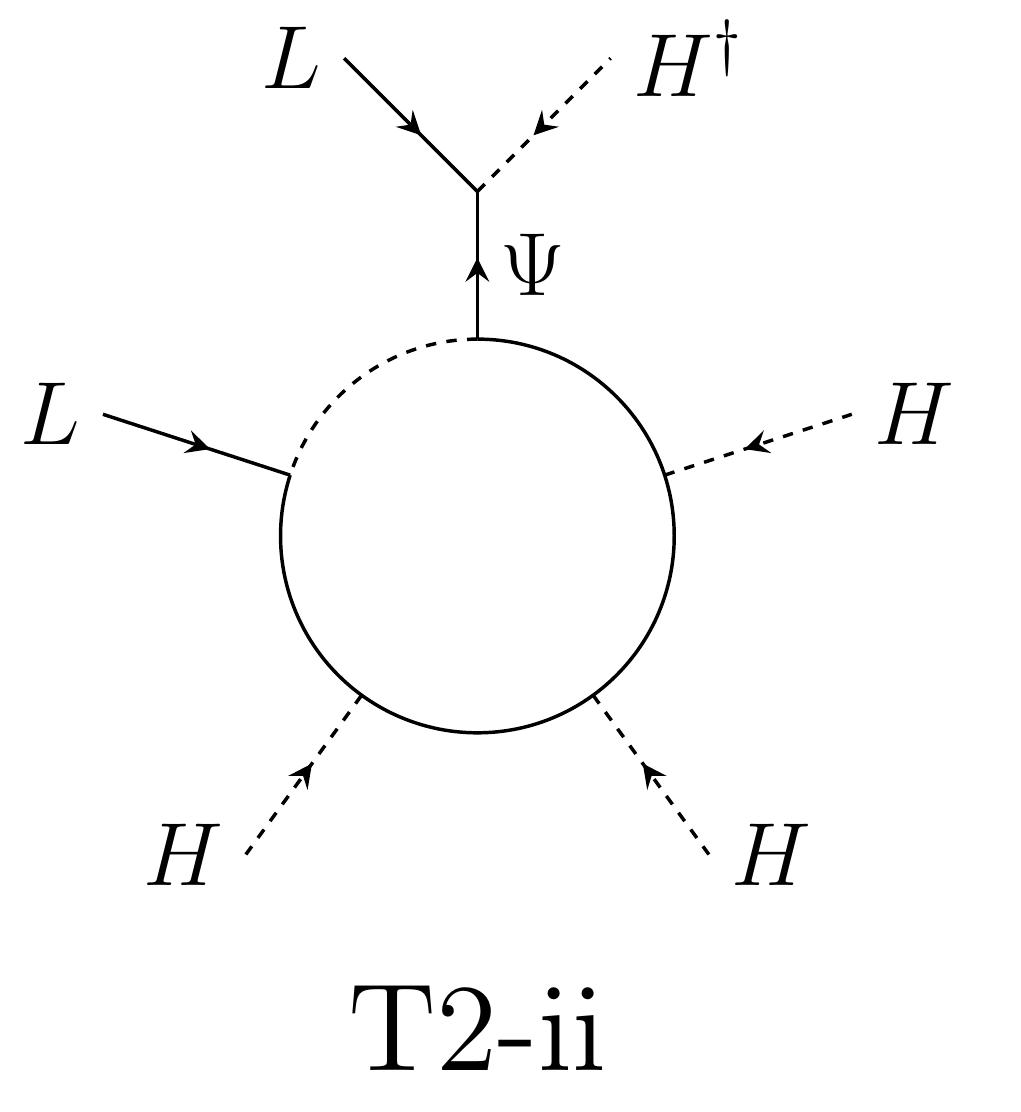}
\end{tabular}
\begin{tabular}{ c c c }
\includegraphics[scale=0.35]{./figures/diagram_T10-i.pdf} \quad & \quad 
\includegraphics[scale=0.35]{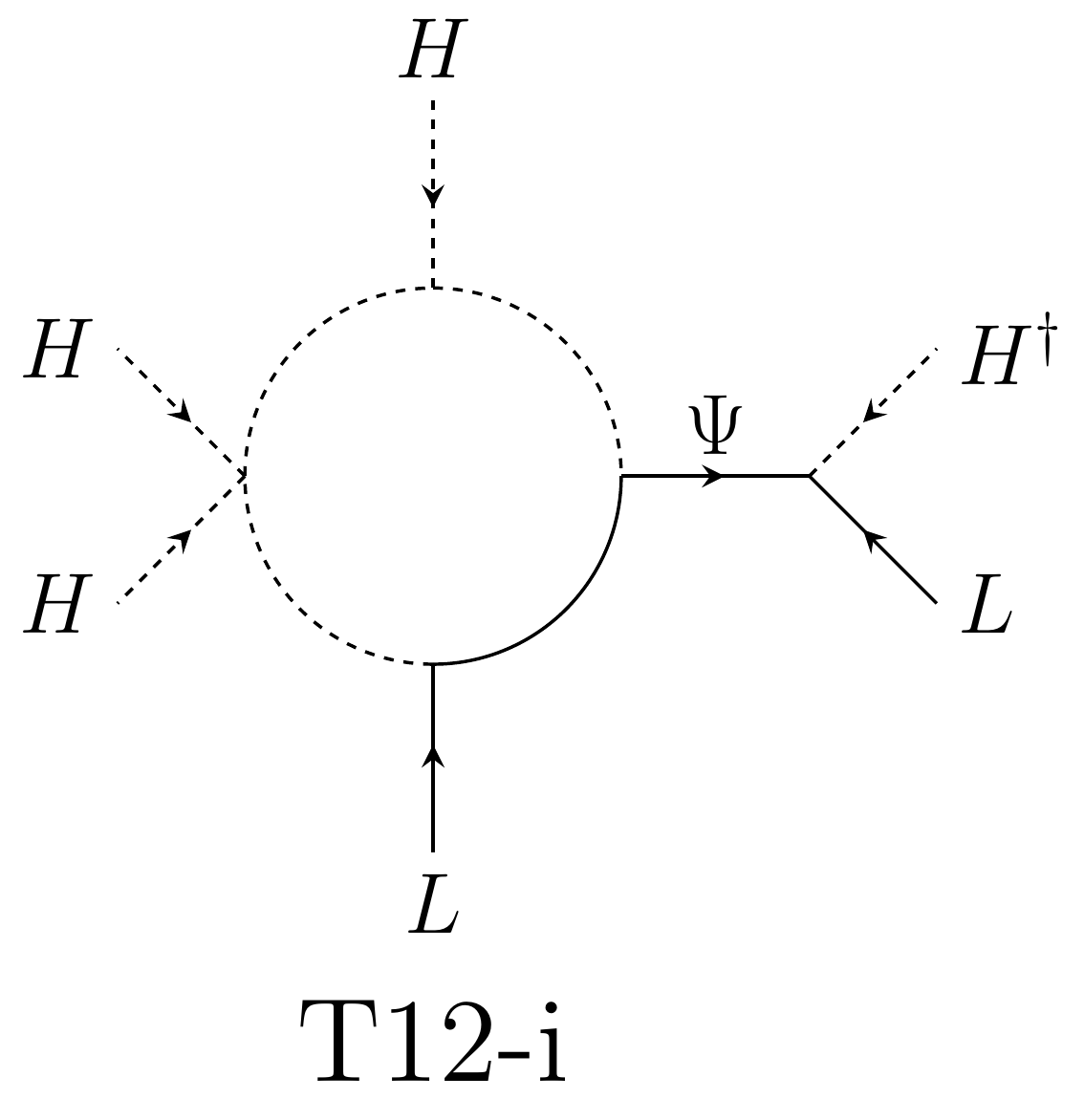} \quad & \quad 
\includegraphics[scale=0.35]{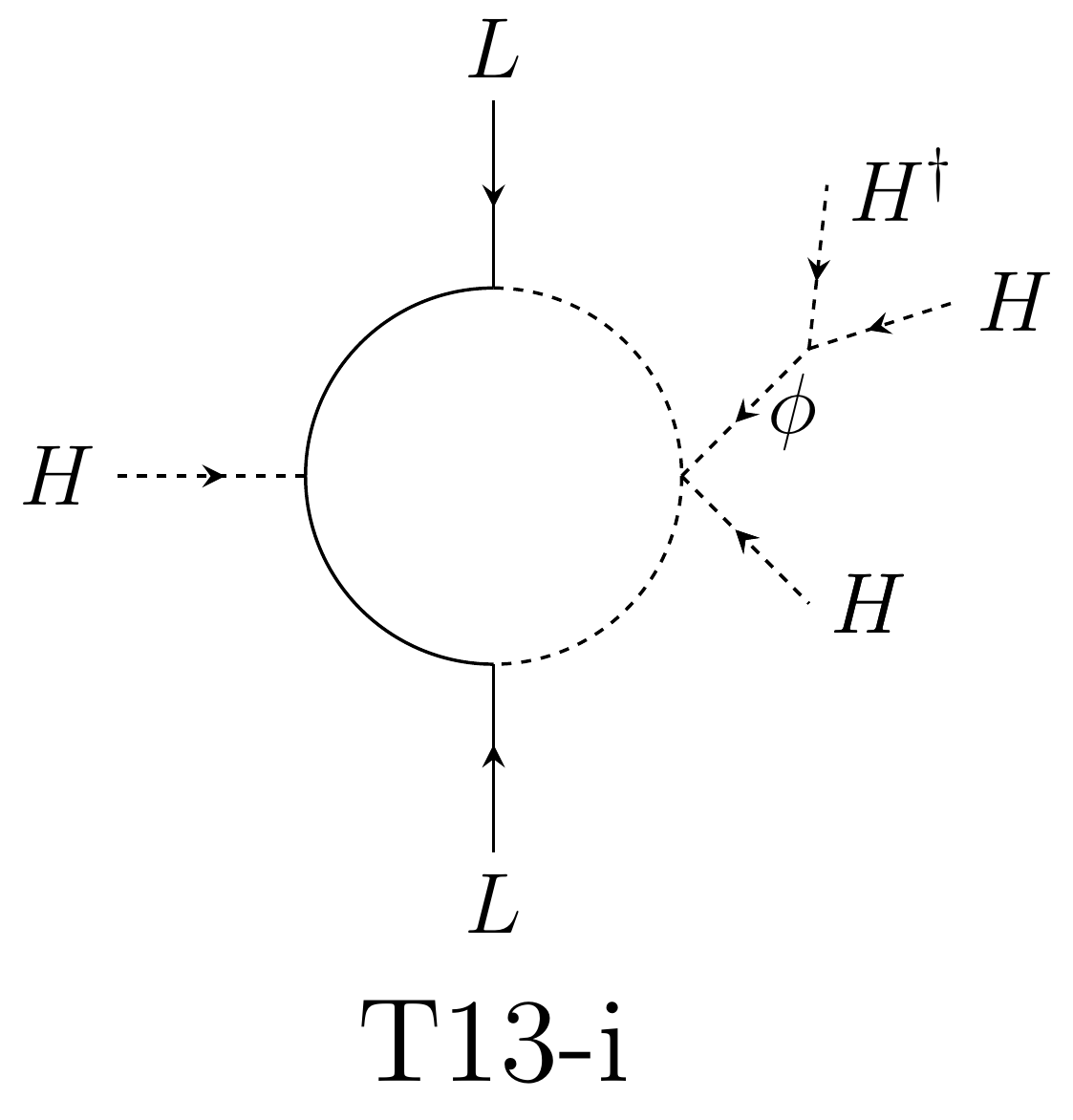}
\end{tabular}
\caption{Diagrams that can lead to a genuine $d=7$ 1-loop neutrino
  mass for which the largest representations of $SU(2)_L$ is at least a
  quadruplet. This group of diagrams require the quadruplet to be one
  of the particles inside the loop to avoid lower order
  contributions.}
 \label{fig:Diags4pletsIn}
\end{figure}

\begin{figure}[h]
\begin{tabular}{ c c c }
\includegraphics[scale=0.35]{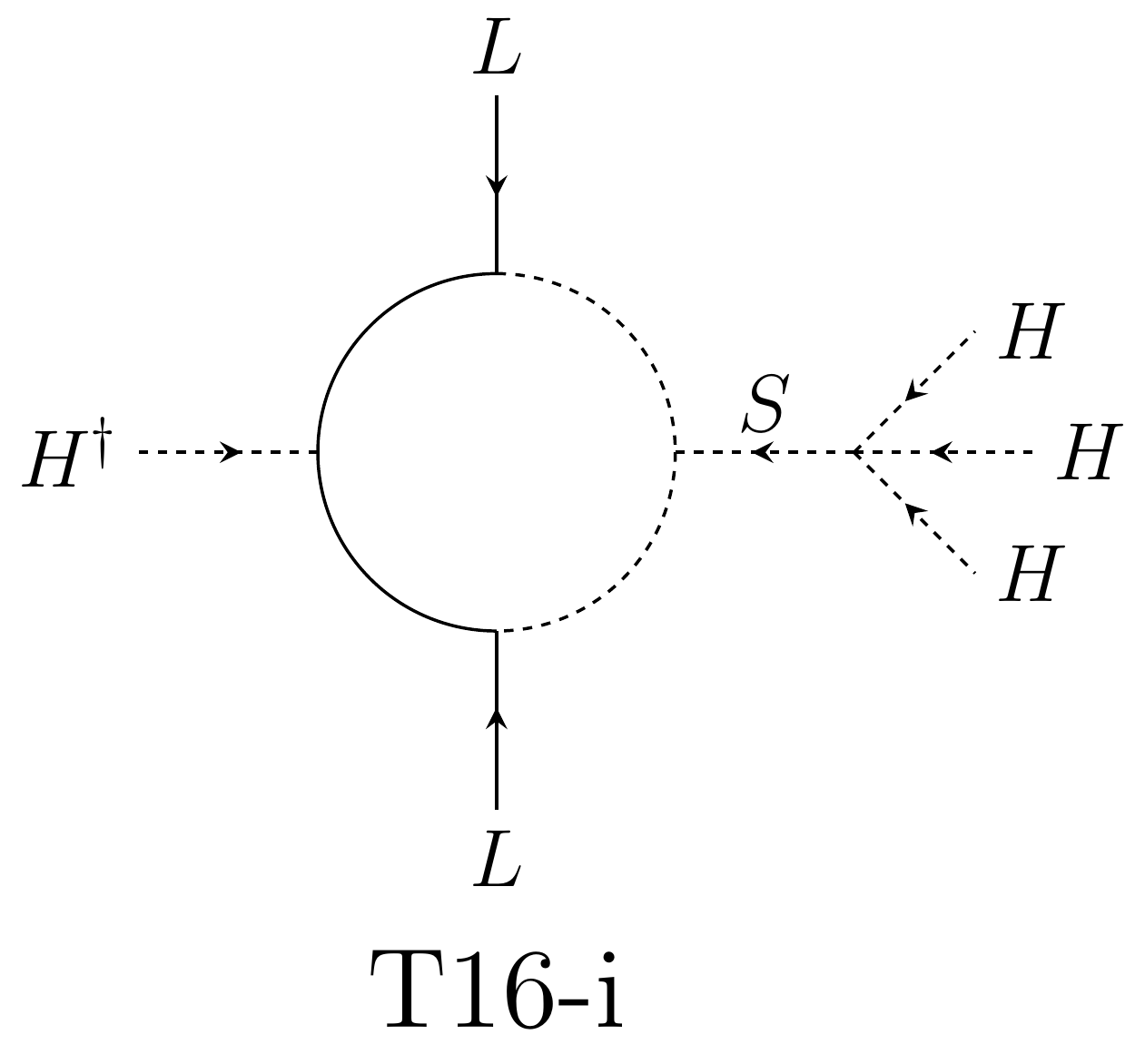} \quad &  \quad 
\includegraphics[scale=0.35]{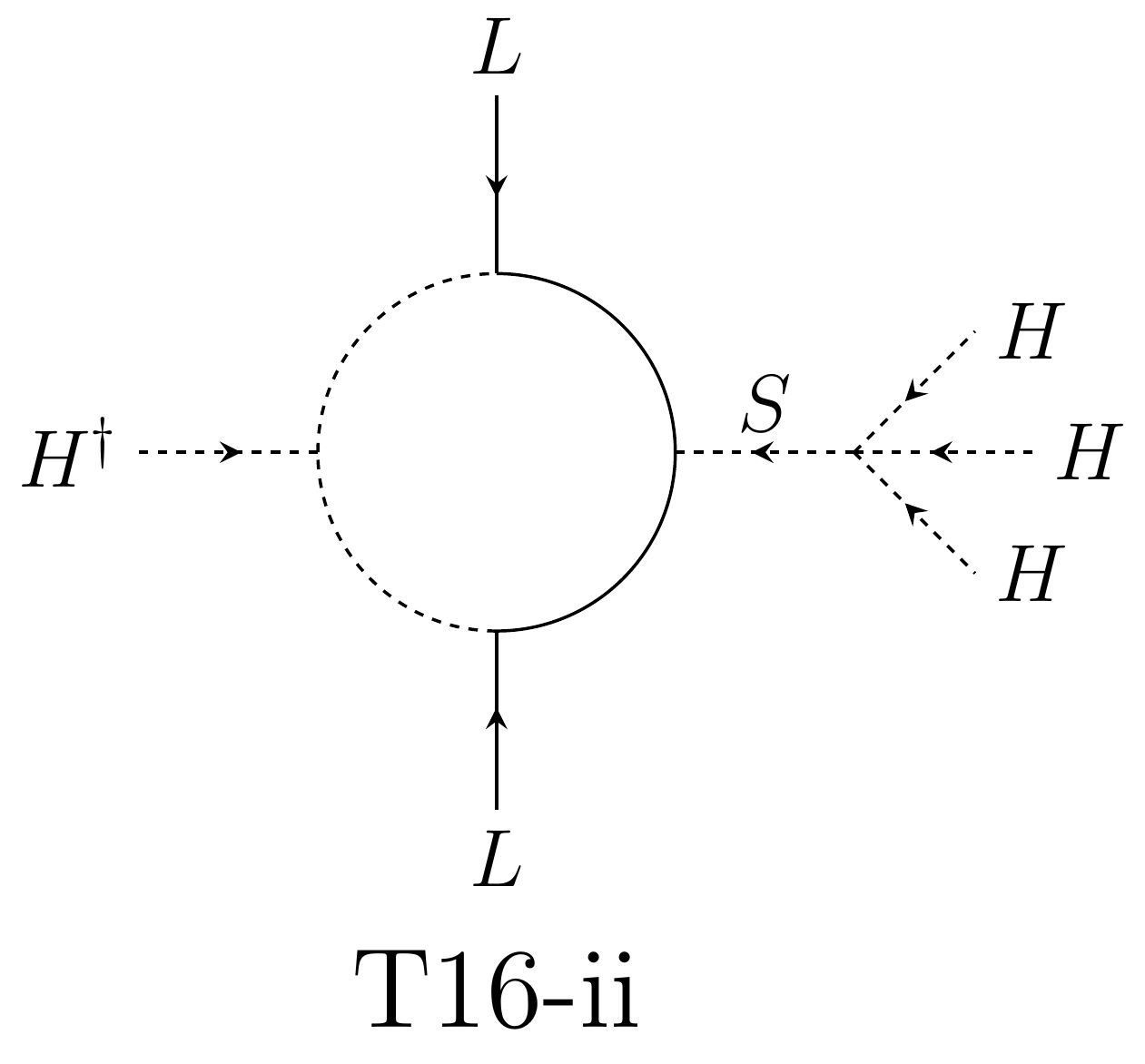} \quad & \quad 
\includegraphics[scale=0.35]{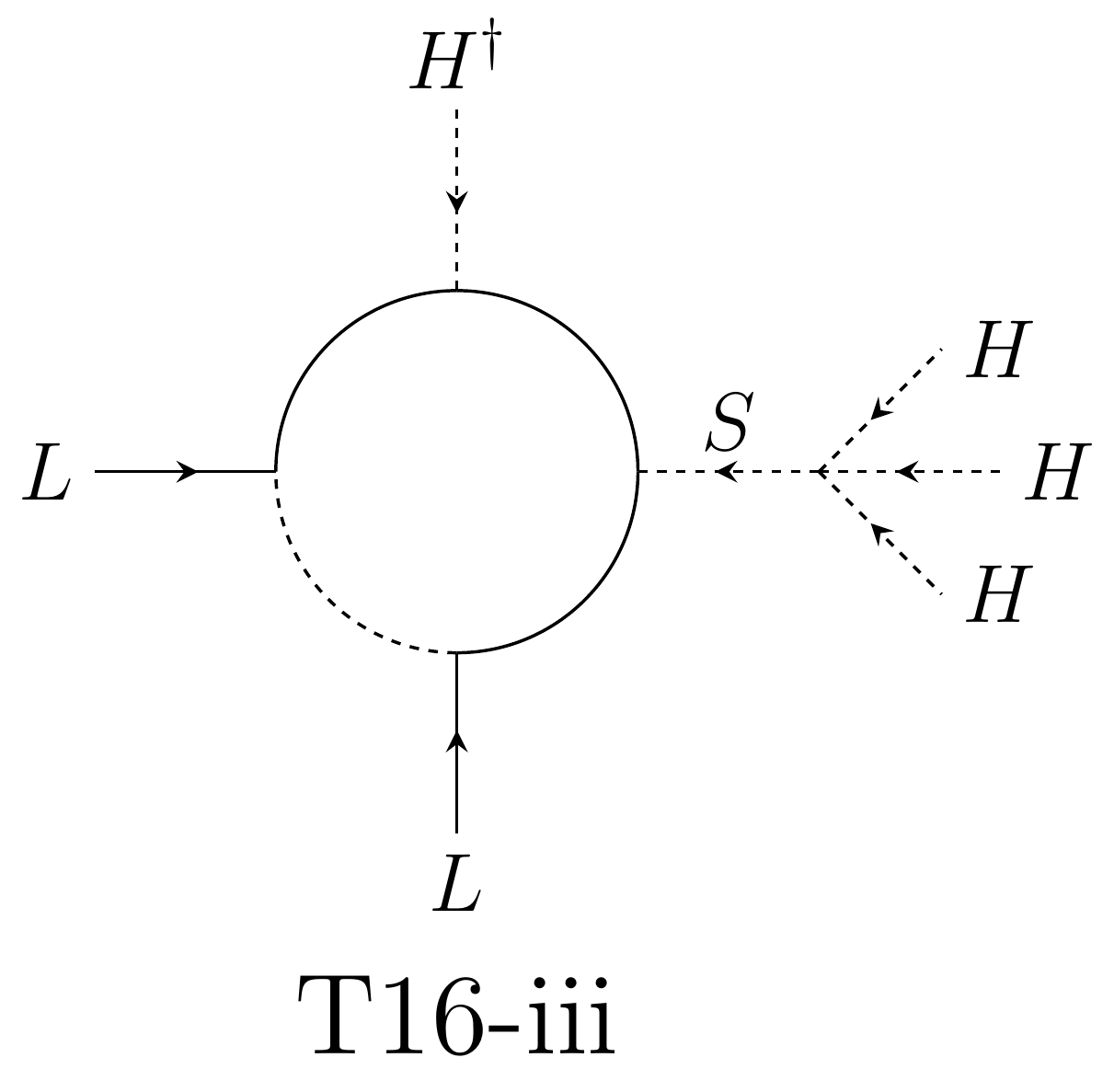}
\end{tabular}
\caption{Diagrams that lead to a genuine $d=7$ 1-loop neutrino mass
  for which the largest representations of $SU(2)_L$ is at least a
  quadruplet. All these diagrams contain $S=\textbf{4}_{3/2}^S$.  The
  hypercharge of the scalar $S$ ensures the absence of a $d=5$ 1-loop
  neutrino mass.}
\label{fig:Diags4pletsS}
\end{figure}

\begin{figure}[h]
\begin{tabular}{ c c c }
\includegraphics[scale=0.35]{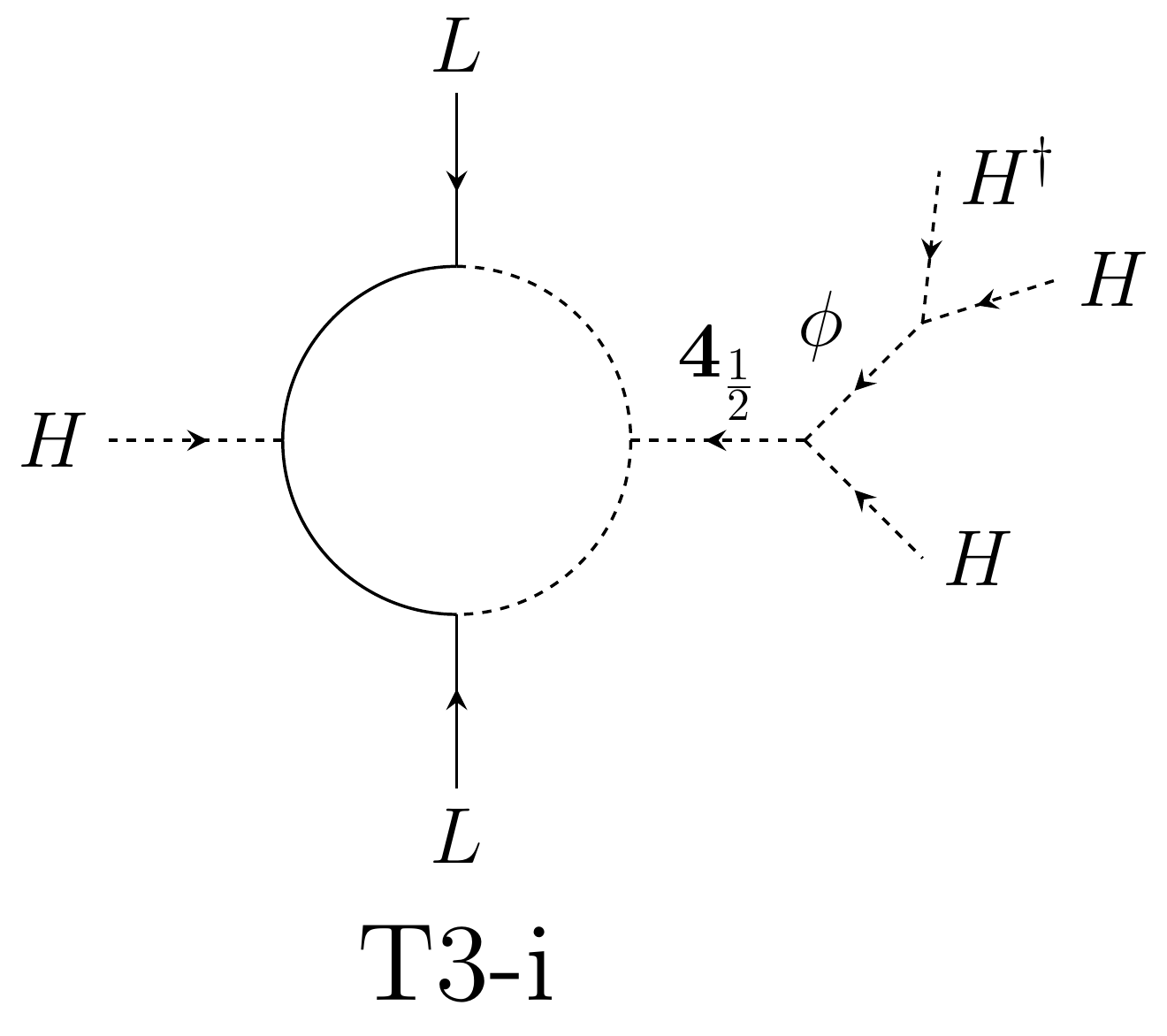} \quad & \quad 
\includegraphics[scale=0.35]{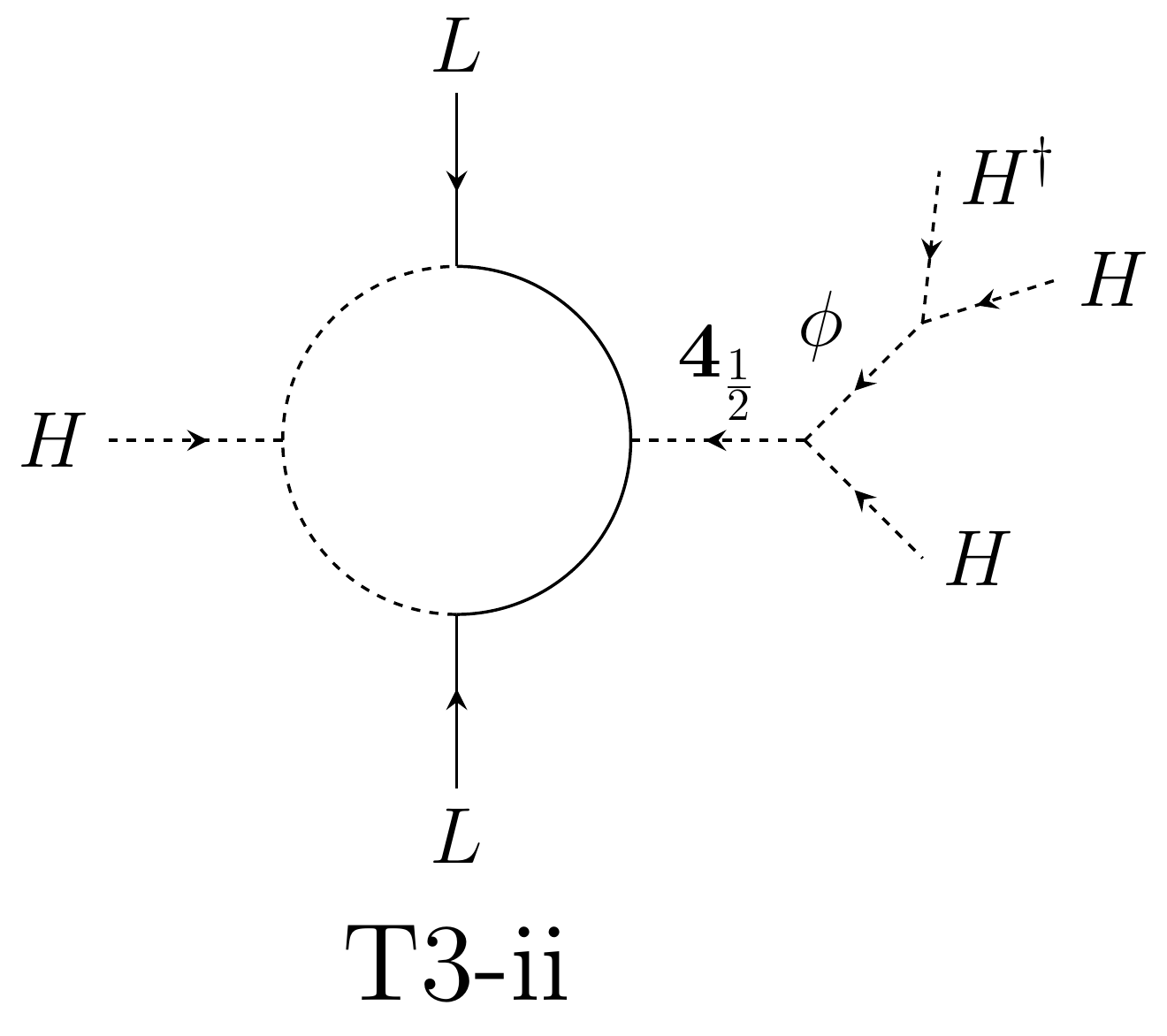} \quad & \quad 
\includegraphics[scale=0.35]{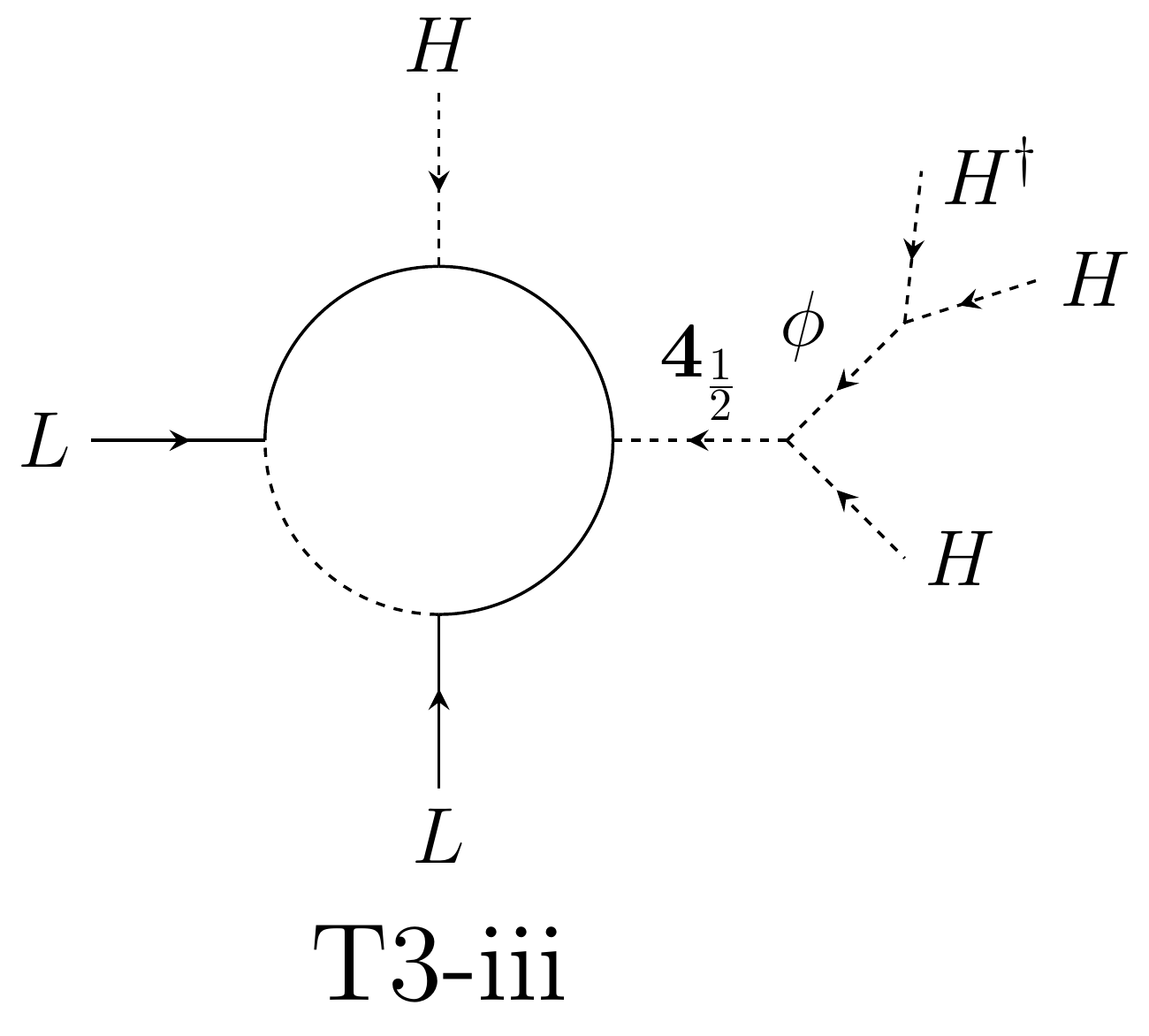}
		 \\
\includegraphics[scale=0.35]{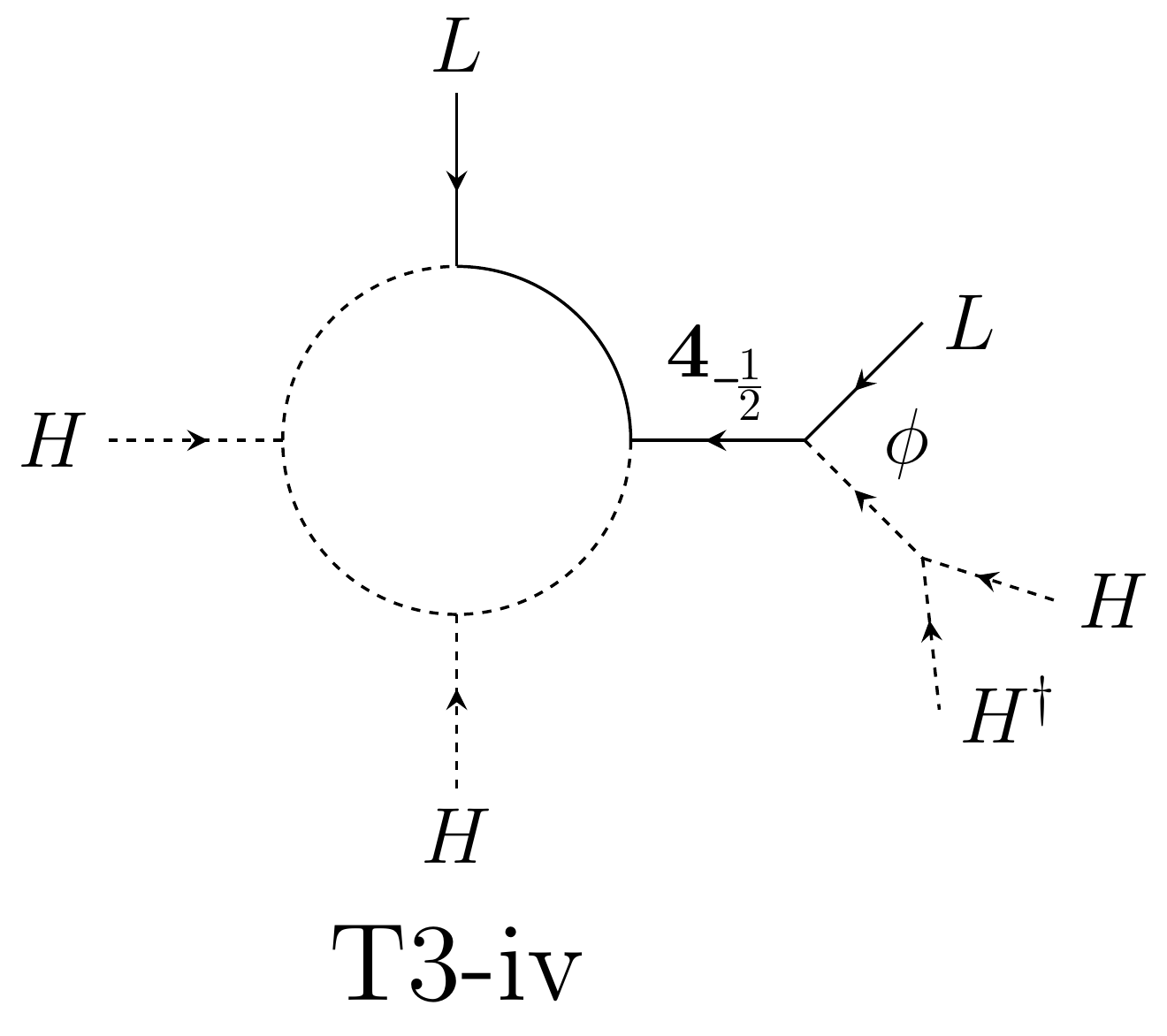} \quad & \quad 	\includegraphics[scale=0.35]{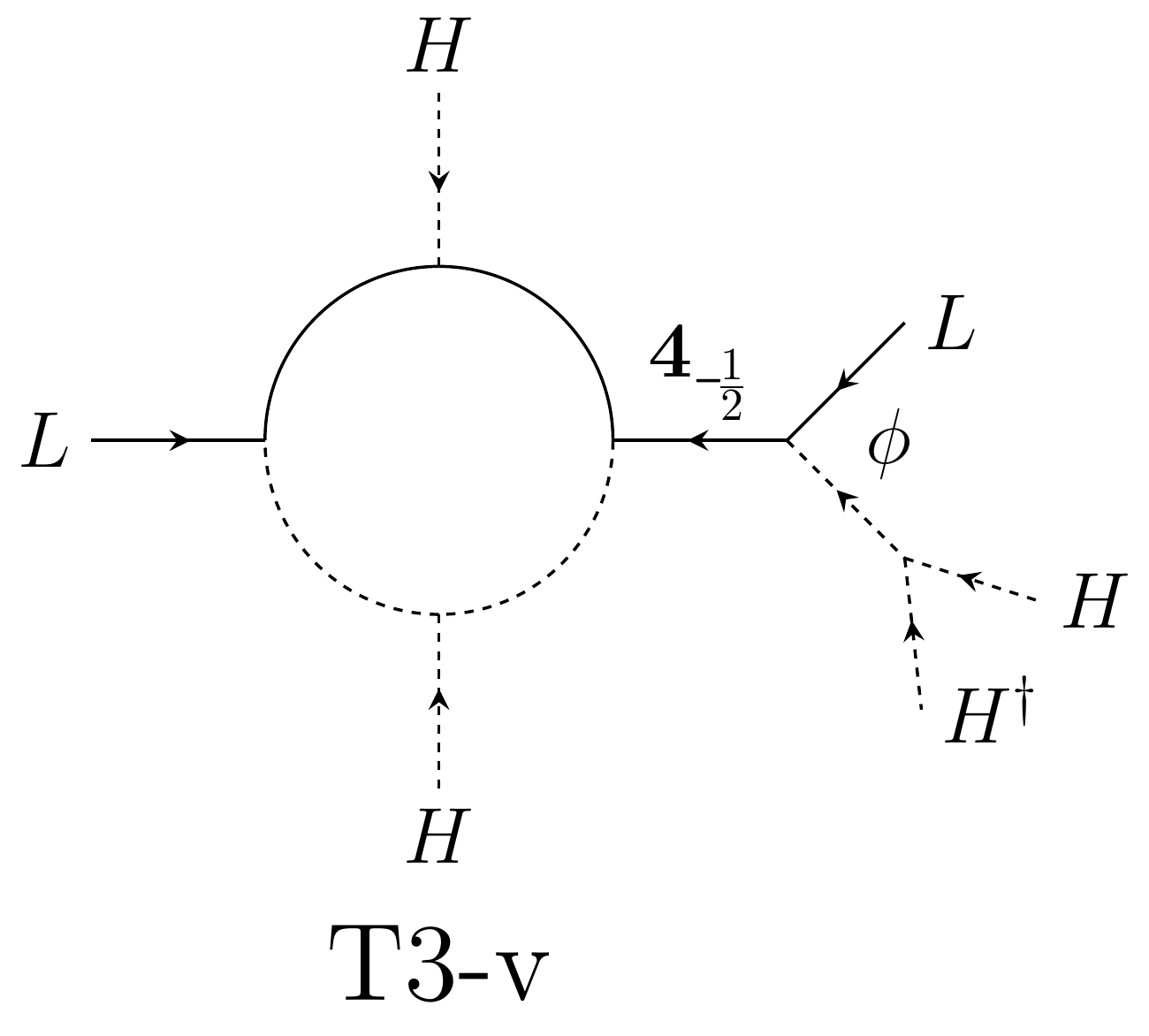} \quad & \quad 
\includegraphics[scale=0.35]{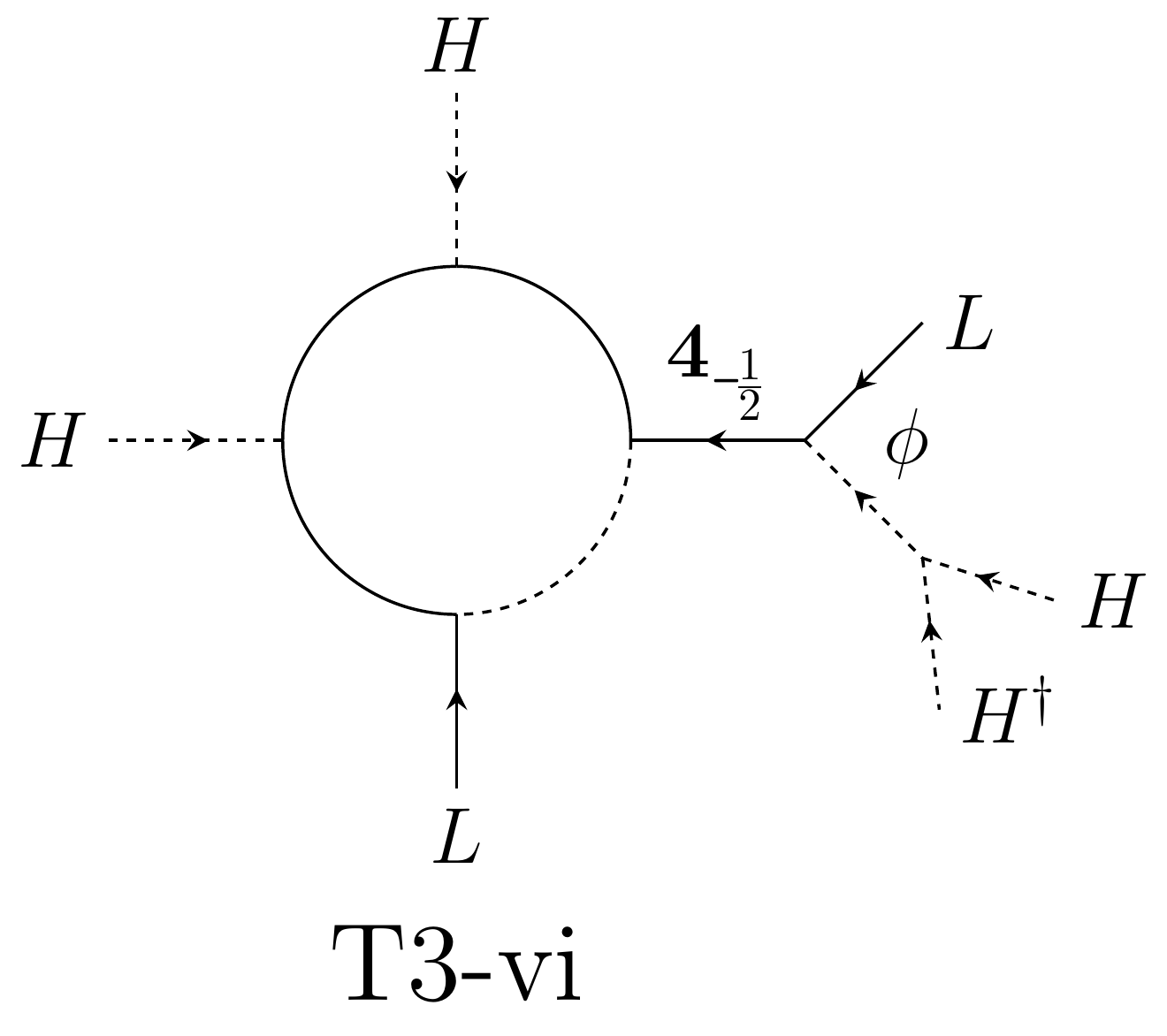}
	 	\\  
\includegraphics[scale=0.35]{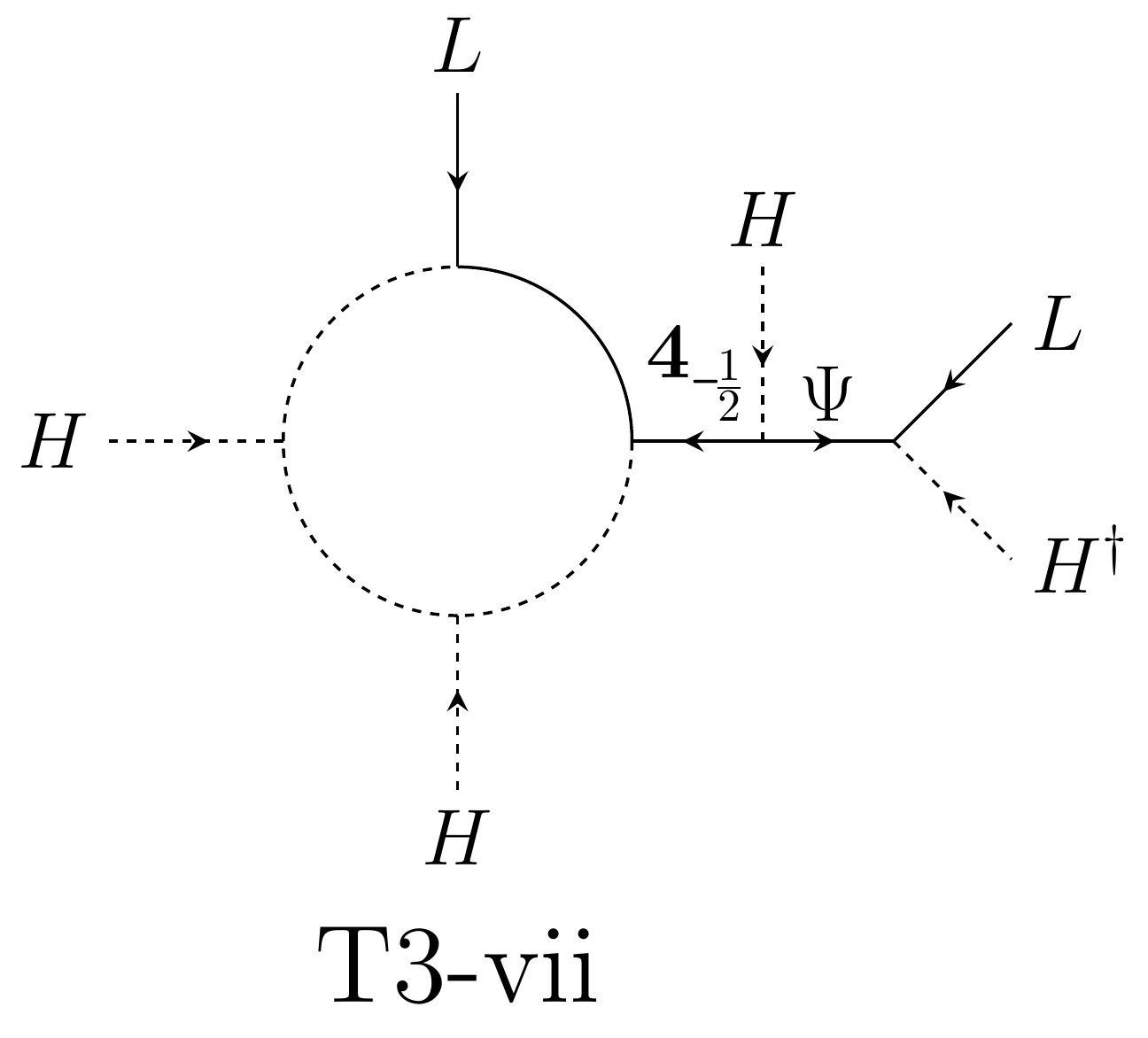} \quad & \quad 
\includegraphics[scale=0.35]{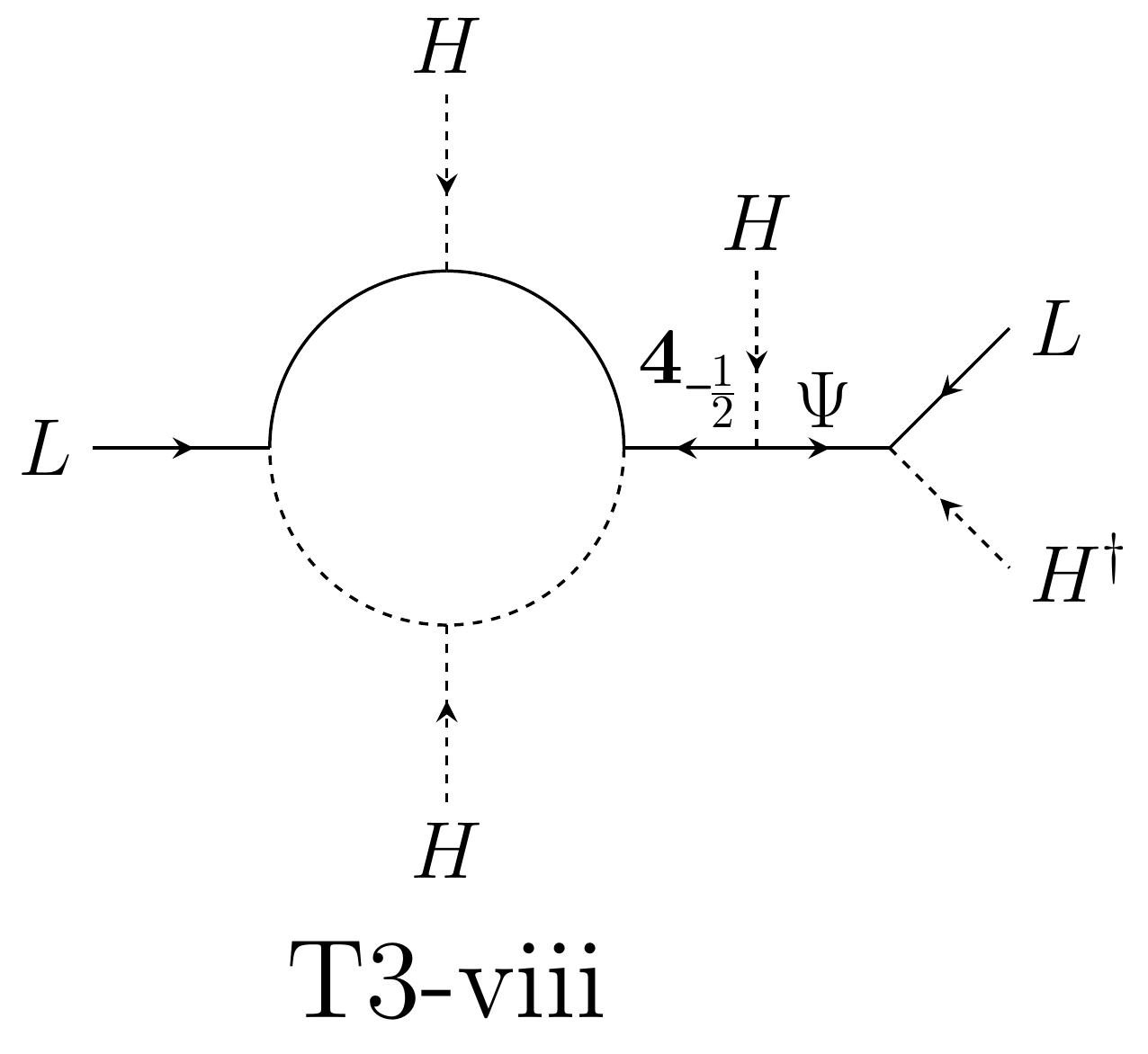} \quad & \quad 
\includegraphics[scale=0.35]{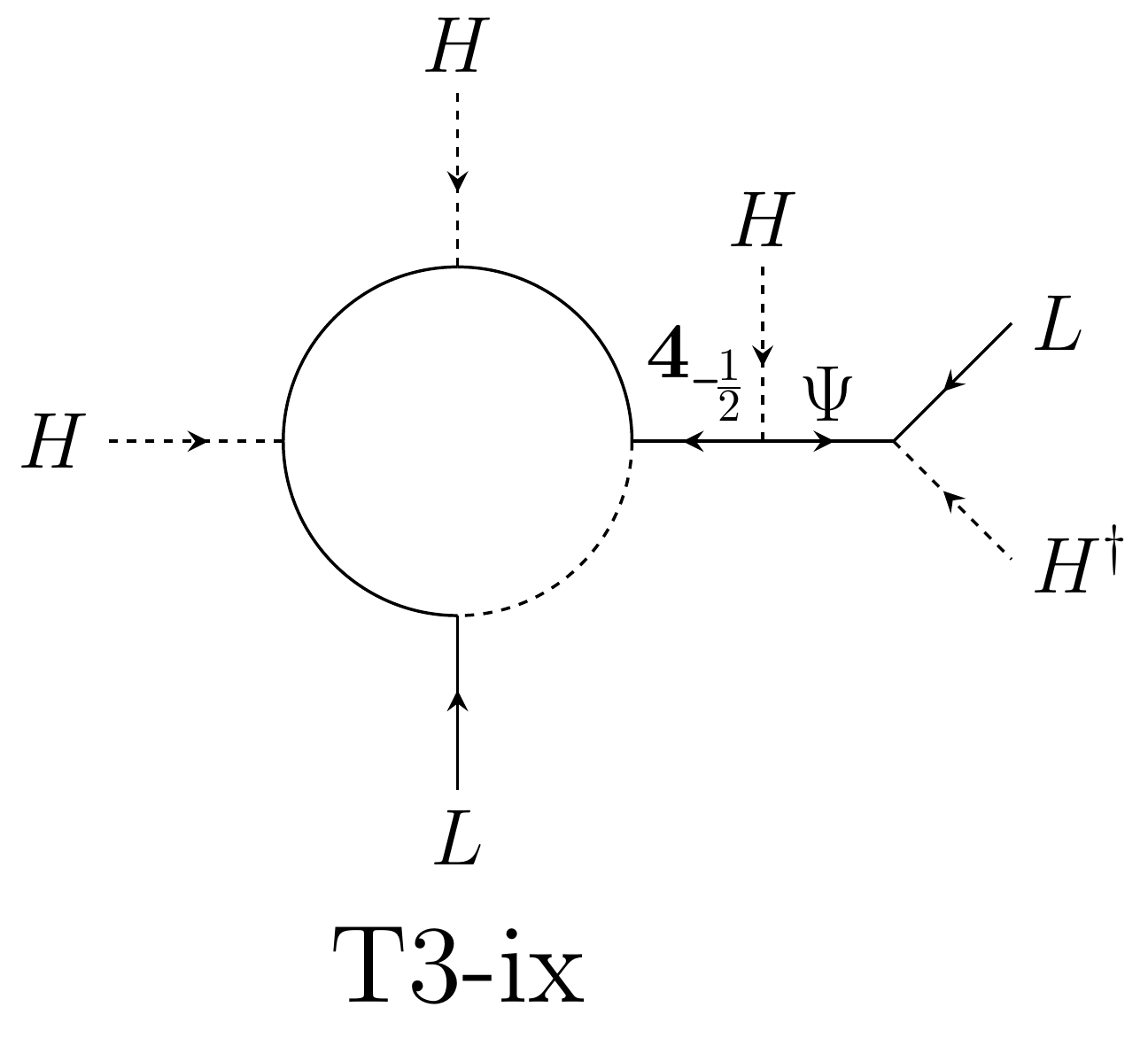}
\end{tabular}
\begin{tabular}{ c c }
\includegraphics[scale=0.35]{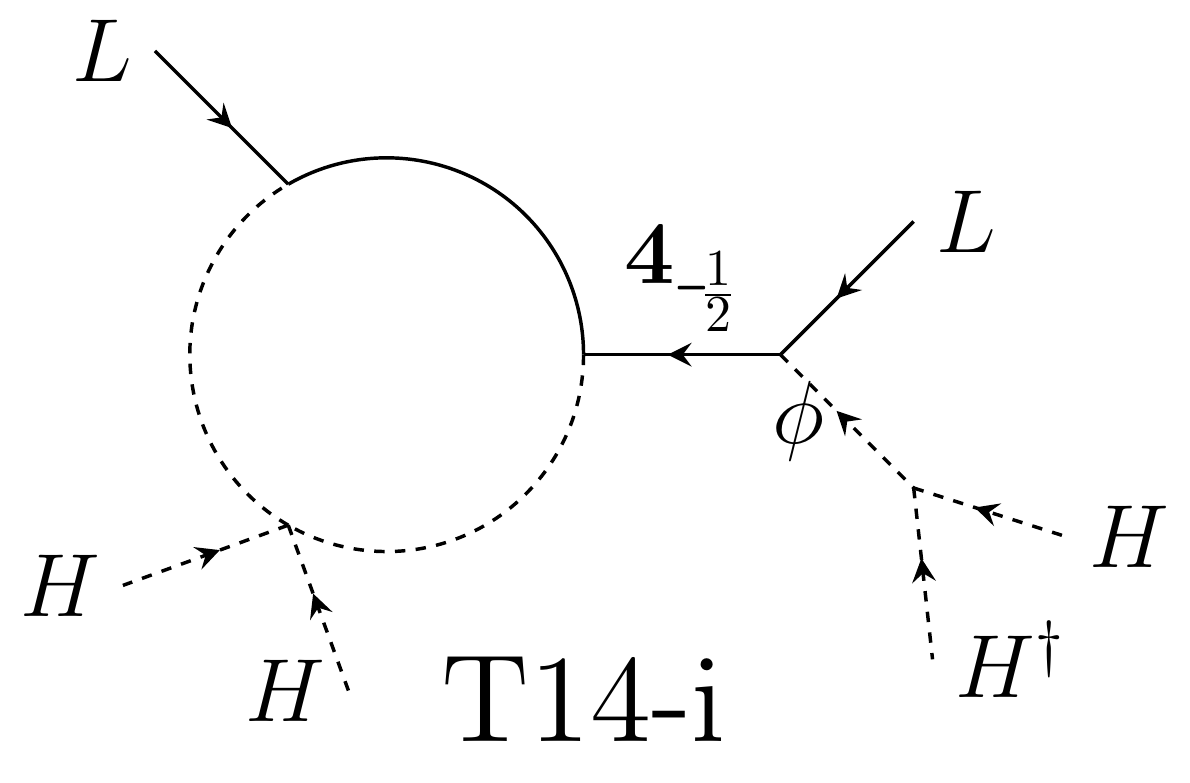} \quad & \quad 
\includegraphics[scale=0.35]{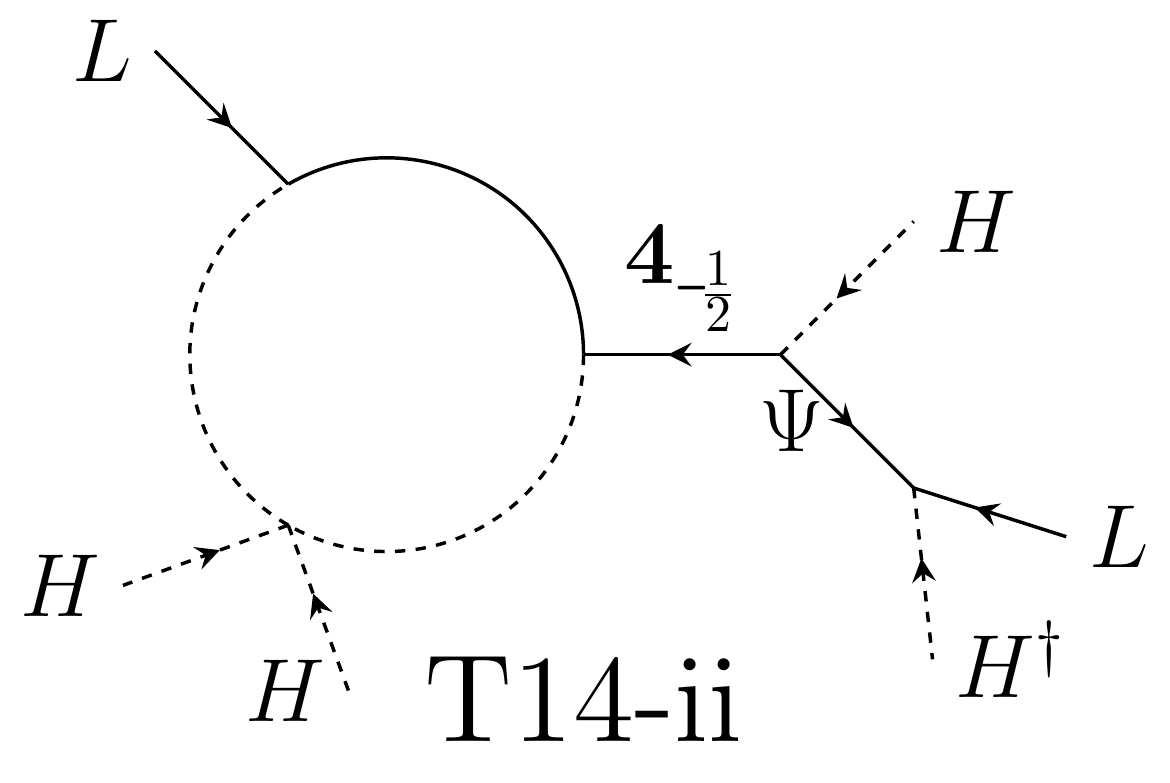}
\end{tabular} 
\begin{tabular}{ c c c }
\includegraphics[scale=0.35]{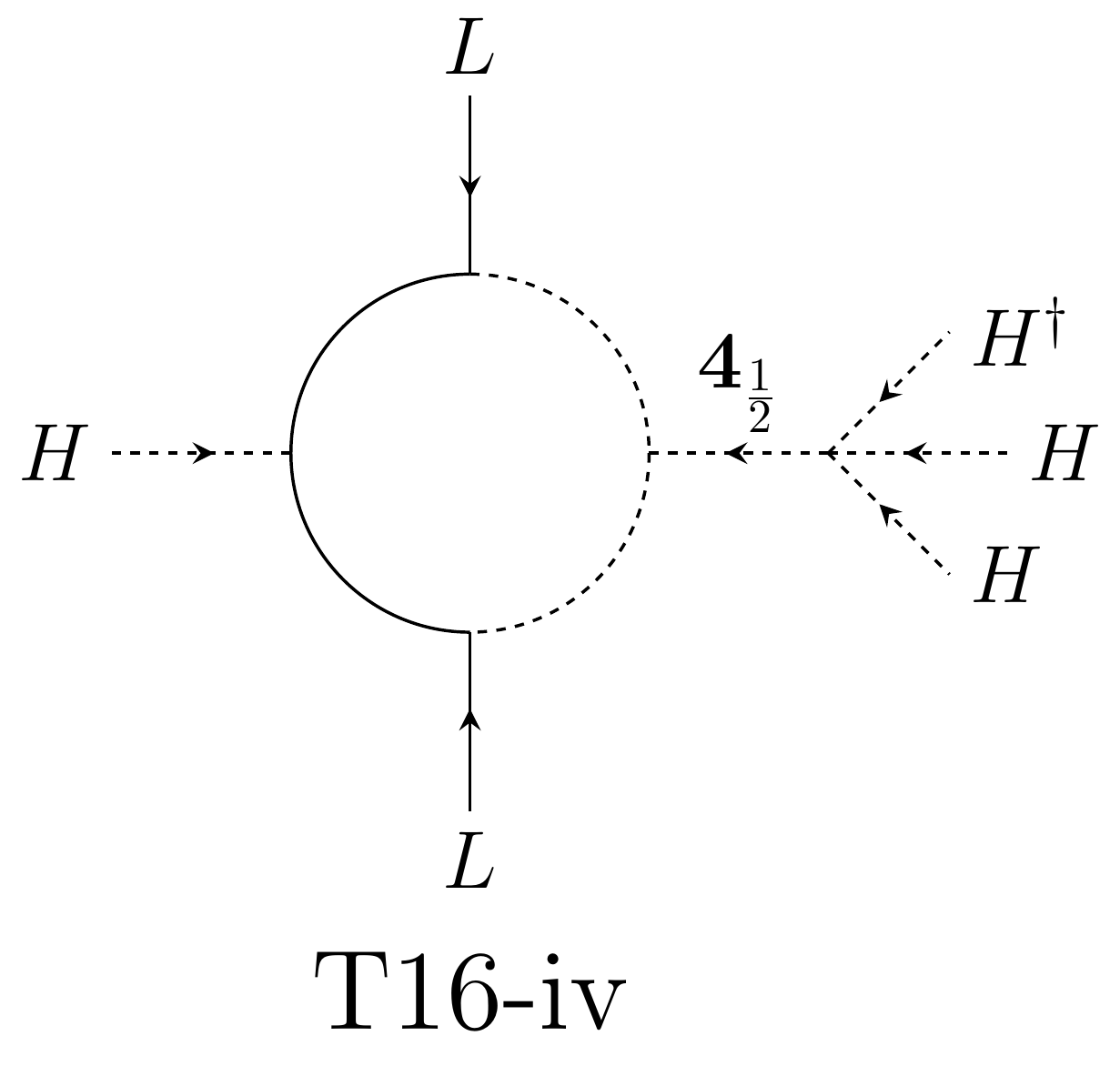} \quad & \quad 
\includegraphics[scale=0.35]{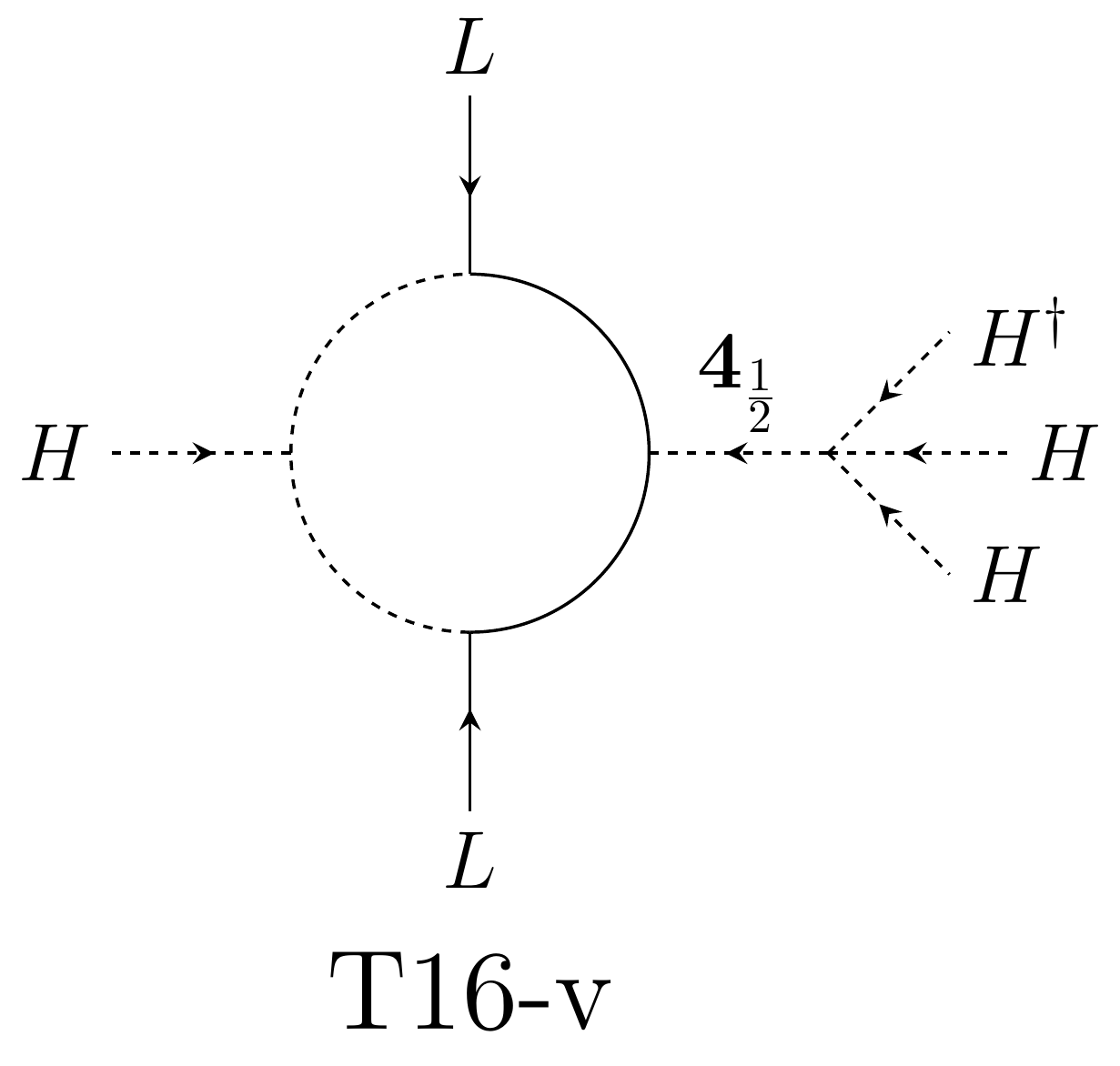} \quad & \quad 
\includegraphics[scale=0.35]{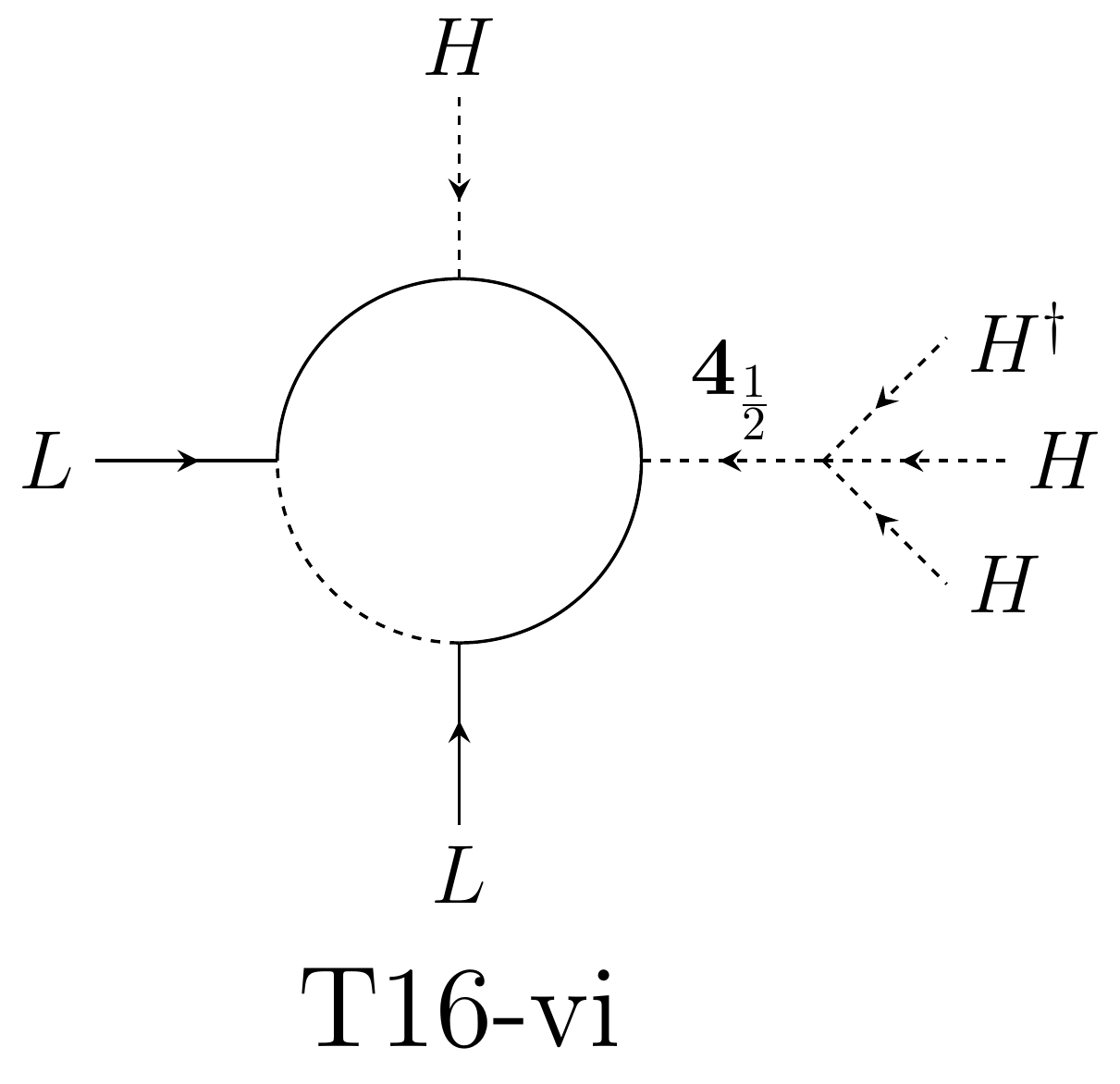}	
\end{tabular}	 
\caption{All remaining diagrams that lead to a genuine $d=7$ 1-loop
  neutrino mass for which the maximum representations of $SU(2)_L$ is at
  least a quadruplet. In these diagrams, two quadruplets are
  needed. Along with the external fermion or scalar quadruplet, a
  genuine models needs an internal quadruplet to allow to distinguish 
  between a $\textbf{4}^S_{1/2}$ ($\textbf{4}^F_{-1/2}$) and a Higgs ($L$). }
\label{fig:Diags4pletsInOut}
\end{figure}

In this section we list diagrams with quadruplets. All diagrams
are given in figs \ref{fig:Diags4pletsIn}, \ref{fig:Diags4pletsS} and
\ref{fig:Diags4pletsInOut}. We have divided these diagrams into
three groups, depending on whether there is a quadruplet in the
loop (fig. \ref{fig:Diags4pletsIn}), the scalar $S$ on the outside
of the loop (fig.\ref{fig:Diags4pletsS}) or models with at least
two different quadruplets (fig. \ref{fig:Diags4pletsInOut}). 

\bibliography{references_0nubb}
\bibliographystyle{h-physrev5}

\end{document}